\documentclass[twocolumn]{aastex7}
\usepackage{subcaption} 
\usepackage{amsmath}
\usepackage{float}
\newcommand{\Ms}{M$_{\odot}$}

\usepackage[normalem]{ulem}
\usepackage{soul}
\shorttitle{Massive star and CSM }
\shortauthors{Sengupta, Sujit \& Sarangi}

\begin{document}
\title{Dance to Demise -- How Massive Stars May Form Dense Circumstellar Shells before Explosion}

\author[0000-0003-0191-4157]{Sutirtha Sengupta}
\affiliation{Indian Institute of Astrophysics, 
100 Feet Road, Koramangala, Bengaluru, Karnataka 560034, India} 
\email{sutirtha.sg86@gmail.com}

\author[0009-0002-3321-4307]{Das Sujit}
\affiliation{Indian Institute of Astrophysics, 
100 Feet Road, Koramangala, Bengaluru, Karnataka 560034, India} 
\affiliation{Pondicherry University, R.V. Nagar, Kalapet, 605014, Puducherry, India}
\email{sujit.das@iiap.res.in}

\correspondingauthor{Arkaprabha Sarangi} 
\author[0000-0002-9820-679X]{Arkaprabha Sarangi}
\affiliation{Indian Institute of Astrophysics, 
100 Feet Road, Koramangala, Bengaluru, Karnataka 560034, India}
\email[show]{arkaprabha.sarangi@iiap.res.in}


\begin{abstract}
We investigate the evolution of red supergiant (RSG) progenitors of core-collapse supernovae (SNe) with initial masses between $12$ and $20$~\Ms, focusing on the effects of enhanced mass loss due to pulsation-driven instabilities in their envelopes and subsequent dynamical ejections during advanced stages of nuclear burning. Using time-dependent mass loss from detailed Modules for Experiments in Stellar Astrophysics (MESA) stellar evolution models, including a parameterized prescription for pulsation-driven superwinds and time-averaged mass loss rates attributed to resulting shock-induced ejections, we construct the circumstellar medium (CSM) before the SN explosion. We calculate resulting CSM density profiles and column densities considering the acceleration of the stellar wind. Our models produce episodes of enhanced mass loss ($\sim 10^{-4}-10^{-2}$~\Ms~$\rm{yr}^{-1}$) in the last centuries-decades before explosion forming dense CSM ($\gtrsim10^{-15}~\rm{g~cm}^{-3}$ at distances $\lesssim10^{15}~\rm{cm}$)  -- consistent with those inferred from multi-wavelength observations of Type II SNe such as SN~2023ixf, SN~2020ywx, SN~2017hcc, SN~2005ip and SN~1998S. The formation of such dense CS shells, within the explored range of our single star RSG models, provides a natural explanation for observed flash-ionization signatures, X-ray and radio emission, and has important implications for dust formation around Type II SNe.
\end{abstract}

\section{Background}
Massive stars with initial masses between $8-30$~\Ms~ \citep{Heger_2003,2003A&A...404..975M} are expected to end their lives in core-collapse (CC) supernovae (SNe), and are considered to be significant dust producers in galaxies \citep{sug06}. However, the evolution of massive stars in their late phases is poorly understood - in particular, mass loss and its impact on the final fate of these stars remains ill-constrained \citep{2014ARA&A..52..487S,2015A&A...575A..60M}. A number of factors have been invoked to explain mass loss from massive stars like line-driven winds, stellar rotation, radial pulsations, episodic ejections/eruptions and binarity \citep{2002RvMP...74.1015W,2012ARA&A..50..107L}. All these factors result in the observed diversity of CCSNe, which are broadly classified as Type I (b/c) -- which do not show any hydrogen lines in their spectra -- and Type II (P/L/n/b) which do so, but with varying strengths and time evolution of light curves \citep{1997ARA&A..35..309F,2022A&A...660A..40M,2025A&A...695A..29S}.

Observationally, direct identifications of SN progenitors \citep{smartt_2009} provide critical information regarding the final stage of its evolution before explosion, and there exist many pre-SN images of Type II progenitors \citep{2015PASA...32...16S,2017hsn..book..693V} The wealth of observations of SNe from large-scale transient surveys like the Zwicky Transient Facility \citep[ZTF;][]{2019PASP..131a8002B}, the Asteroid Terrestrial-impact Last Alert System \citep[ATLAS;][]{2018PASP..130f4505T} and the upcoming Legacy Survey of Space and Time \citep[LSST;][]{2023PASP..135j5002H}, provides golden opportunities to improve our current understanding of the final stages leading to the explosive end of the lives of massive stars \citep{2024arXiv240910596R,2024ApJ...960...72S}.

However, several factors complicate the task of deriving reliable mass loss rates ($\dot M$) for cool, possibly dust-enshrouded, red supergiant (RSG) progenitors of Type II SNe \citep{2005A&A...438..273V,2021ARA&A..59..337D}. At the same time, there is a growing consensus that a high fraction of Type II SNe are preceded by phases of heavy mass loss \citep{10.1093/mnras/staf888} whose signatures can be seen from the post-explosion shock interaction with the circumstellar medium (CSM) created by the wind of the SN progenitor \citep{2018SSRv..214...27C,2024arXiv240504259D,universe11050161}. The circumstellar densities inferred from flash-ionization spectroscopy, X-ray and radio observations of SNe like SN~1998S \citep{10.1046/j.1365-8711.2002.05086.x}, SN~2010jl \citep{2014ApJ...797..118F}, and SN~2023ixf \citep{2023ApJ...956L...5B} point toward mass loss rates of $\sim 10^{-3}-10^{-1}$~\Ms~$\rm{yr}^{-1}$ within $\sim 10^{15}-10^{16}$~cm of the progenitor —- values that far exceed those from standard steady winds. The presence of narrow P~Cygni absorption features, light curve plateaus with CSM interaction signatures, and late-time IR emission all corroborate the scenario of enhanced pre-SN mass loss. All these observations of post-explosion features are clearly at odds with latest mass loss prescriptions inferred from large samples of observed RSGs \citep{10.1093/mnras/stad1818,2023ApJ...942...69M,2023A&A...676A..84Y,antoniadis_2024,2024A&A...681A..17D} as well as empirical mass loss rates derived from much smaller samples of stars \citep{1988A&AS...72..259D,1990A&A...231..134N} that are still used in state-of-the-art stellar evolution codes to model the RSG phase of massive stars \citep{2011A&A...526A.156M,2021ApJ...922..177M}.

From a theoretical point of view, there have only been handful of attempts to derive mass loss prescriptions for RSGs from first principles. \cite{2010ApJ...717L..62Y} proposed a pulsation-driven wind mass loss prescription for high mass RSGs with high luminosity-to-mass ratios, which they hypothesized could, in principle, achieve the high rates ($\dot M \sim 10^{-2}$~\Ms~$\rm{yr}^{-1}$) required to explain observed properties of CSM around Type IIn SNe. Thereafter, \cite{ml2015} followed similar principles to propose a pulsation-driven wind for very massive metal-free stars. \cite{clayton2018a} further extended the pulsation-driven mass loss into the shock-dominated regime and proposed analytical mass loss rates for use in stellar evolutionary calculations. \cite{2021A&A...646A.180K} proposed an alternate mechanism to drive mass loss through the pressure due to characteristic turbulent velocities inferred for RSGs. Most recently, \cite{2023A&A...678L...3V} proposed a new mass loss prescription for RSGs as a function of the Eddington factor based on the data from \cite{2023A&A...676A..84Y}. However, none of these existing prescriptions by itself can fully explain all the observational constraints such as the RSG luminosity functions \citep{2023ApJ...942...69M}. Even binary interactions are estimated to overpredict luminous RSGs arising from mass accretion or mergers  \citep{2025A&A...697A.167Z}. Hence, most evolutionary codes continue to employ empirical mass loss rates \citep{2021ApJ...922...55B} which lack physical grounding and fail to reconcile with observed properties of Type II SNe.

\section{Rationale of this study}

The motivation behind this work is to address (a) whether a single star may incur heavy mass loss in the final stages before it explodes, that can be correlated with the presence of a dense CSM near the star, and (b) how the progenitor mass impacts the mass loss rates, CSM densities, and wind velocities. In this regard, our study accounts for progenitors of Type II supernovae, and we compare the CSM profiles of well-observed interacting SNe such as SN~2023ixf \citep{2023ApJ...952L..30J,2023ApJ...954L..42J, 2024Natur.627..759Z,2023ApJ...956...46S,2024ApJ...973L..47B, 2025ApJ...985...51A}, SN~2005ip \citep{Smith_2009,Katsuda_2014}, SN~1998S \citep{10.1046/j.1365-8711.2002.05086.x,2004MNRAS.352..457P,Fransson_2005,2012MNRAS.424.2659M,2015ApJ...806..213S}, SN~2017hcc \citep{2022MNRAS.517.4151C}, SN~2020ywx \citep{Baer-Way_2025}.  
We argue that for single stars, enhanced pulsation near the end of their lives provides conditions favorable for escalated mass loss \citep{2010ApJ...717L..62Y,ml2015}. To demonstrate this effect, we simulate the evolution of massive stars with initial masses between $12$ and $20$~\Ms, from their main-sequence until the end of neon burning (or until the end of core O-burning, in some cases) using the MESA code, applying advanced mass loss prescriptions to incorporate effects of radial pulsations in the outer envelope of RSGs (Section~\ref{section: pulsating RSGs}).

The paper is arranged as follows: in Section \ref{section: mass loss}, we briefly review the various mass loss mechanisms in RSG progenitors of CCSNe along with observational evidence for enhanced mass loss. In Section \ref{section: pulsating RSGs}, we describe our modeling approach for incorporating enhanced mass loss in pulsating RSG progenitors followed by the model predictions for the properties of the CSM. In Section \ref{section : comparison}, we compare our model predictions with available observational estimates of mass loss rates, CSM density and column density along line of sight and discuss the implications of our model predictions for upcoming infrared observations from large surveys like the James Webb Space Telescope \citep[JWST;][]{Gardner_2023} and the LSST in Section \ref{section: discussions}. Finally, we conclude with future directions of work in synergy with upcoming transient surveys.

\section{Mass loss in RSG\MakeLowercase{s}}\label{section: mass loss}

Mass loss from RSGs is a fundamental yet still uncertain ingredient in the late stages of stellar evolution \citep{10.1088/978-0-7503-1278-3ch15,galaxies13020025,galaxies13020033,galaxies13040072,galaxies13040081}. The physical mechanisms responsible for this process directly shape the CSM, influencing both the pre-SN environment and the post-explosion observables, including light curves, spectra, and dust formation \citep{10.1088/978-0-7503-1329-2ch4}. Observations of Type II SNe with dense CSM (e.g., SN~2009ip, SN~2010jl, SN~2023ixf and so on) have revealed that enhanced and often eruptive mass loss can occur shortly before CC \citep{2012AJ....144..131Z,2013ApJ...768...47O, 2023ApJ...956L...5B}.
In this section, we describe potential driving mechanisms for steady line-driven winds, pulsation-driven outflows including dynamical ejections and/or eruptions, turbulent pressure and convective shocks creating extended atmosphere/chromosphere for dust-driven winds, and enhanced mass loss due to binary interaction for RSGs.

\subsection{Line-driven wind}

Line-driven winds are driven by the transfer of momentum from photons to metal lines in the stellar atmosphere. This mechanism is efficient in hot, compact stars e.g., OB stars \citep{2017RSPTA.37560269V}, but is considerably less effective in RSGs due to their cooler photospheres ($T_\mathrm{eff} \lesssim 4000$~K), where fewer atoms are ionized and UV photons are scarce \citep{1975ApJ...195..157C, 1997A&AS..123..305V}. Typical mass loss rates from line-driven winds in RSGs range from $\sim 10^{-7}$ to $10^{-6}$~\Ms~$\rm{yr}^{-1}$ \citep{2021ApJ...922...55B,2025arXiv250717052M}, which is inadequate to explain the dense CSM observed in the majority (upto $80\%$) of all Type II SNe with early-time observations \citep{2018ApJ...858...15M,2018NatAst2.10.808}.

\subsection{Pulsation driven mass loss}

Many RSGs exhibit large-amplitude radial pulsations with periods of several hundred days \citep{2018ApJ...859...73S, 2019ApJS..241...35R}, as observed in long-period variables (LPVs). Such radial motions in the H-rich envelope of RSGs are driven by the so-called $\kappa$-mechanism \citep{1965ApJ...142..868B, 2008A&A...484...29G} - an interplay between excess radiation pressure and gravitational pull on the outer layers of the RSG envelope. These pulsations can grow in amplitude to produce shocks at the stellar surface that levitate material above the photosphere, reducing the effective gravity and allowing other processes (e.g., radiation pressure on dust and molecules) to accelerate the material outward. However, the exact mechanism of mass loss due to the complex interplay of pulsations and convection in RSG envelopes remains ill-understood \citep{2010Ap&SS.328..245G}, and attempts to reduce these inherently three-dimensional phenomena into one-dimensional prescriptions for use in evolutionary codes remain limited \citep{2010ApJ...717L..62Y,ml2015,2025A&A...703A..61B,2025MNRAS.543.3929S}.

\subsubsection{Dynamical Ejections}

In advanced nuclear burning stages, vigorous convection and shell flashes can lead to hydrodynamical instabilities and formation of shock waves in the outer envelopes of RSGs. If these internal shocks propagate outward and reach the stellar surface, they can trigger impulsive ejections of envelope material. These ``dynamical ejections'' are thought to be responsible for the formation of dense, compact shells detected in SN progenitors such as SN~2009ip \citep{2013MNRAS.430.1801M,2014ARA&A..52..487S}. Simulations suggest that mass loss rates in such events can temporarily exceed $10^{-2}$~\Ms~$\rm{yr}^{-1}$, lasting sometimes for days to weeks or even sustain repeatedly for centuries \citep{clayton2018a}.

\subsubsection{Eruptive mass loss}

Eruptive mass loss may occur through surface instabilities driven by opacity changes, pulsationally enhanced convective transport, wave-driven energy deposition or super-Eddington luminosities in RSG envelopes \citep{2012MNRAS.423L..92Q,2017MNRAS.470.1642F,Cheng_2024}. These eruptions resemble luminous blue variable (LBV) behavior, though they are seen in cooler RSGs as well. The pre-SN outbursts seen in light echoes (e.g., SN~2013cu) are likely signatures of such events, with inferred $\dot{M} \sim 10^{-4}$–$10^{-2}$~\Ms~$\rm{yr}^{-1}$ over timescales of months to years \citep{2016MNRAS.455..112G}. Furthermore, statistical analysis of luminous stars in the Milky Way \citep{2011ApJ...743...73K} and the study of $\eta$ Carinae analogs in nearby galaxies \citep{2015ApJ...815L..18K} reveal the presence of massive shells which hint towards instabilities triggered at the onset of C-burning leading to eruptions \citep{Margutti_2017}.

\subsection{Turbulent pressure and Extended Chromosphere}

\cite{2021A&A...646A.180K} have recently presented a model showing how turbulent pressure inside the RSG envelope is sufficient to expand the atmosphere to an extent that a wind can be launched. Their analytic mass loss prescription for observationally inferred turbulent velocities \citep{2015A&A...575A..50A} predicts that stars above $M_{\rm{init}} \gtrsim 17$\Ms~leave the RSG phase to evolve into blue supergiants and hence would not explode as TypeII P/L SNe -- in agreement with observations \citep{2009MNRAS.395.1409S} and in contrast to results obtained using empirical mass loss rates \citep{1988A&AS...72..259D,2021ApJ...922...55B} in which massive stars below $30$\Ms end their lives as RSGs.

Along similar lines, \cite{2024OJAp....7E..47F} proposed the existence of a dense chromosphere supported by outgoing shocks near the photosphere created from strong convective motions inside the RSG envelopes. The chromosphere in their model extends to several stellar radii \footnote{In \cite{2024OJAp....7E..47F}, the stellar radius is the outer boundary of the relaxed stellar model, located at a very low optical depth so that their models give a smooth and realistic transition to the circumstellar environment.}, out to distances where dust formation can drive the wind launched by the radiation pressure of the star \citep{1995SSRv...73..211S}. The mass loss rates predicted from their model are similar to recent observations, with increasing values for more luminous RSGs, and can potentially explain the early rise in light curves of Type IIP SNe without invoking pre-explosion enhanced mass loss.

\subsection{Dust driven wind}\label{dust driven wind}

Beyond a few stellar radii, the temperature in RSG atmospheres drops below $\sim 1500$~K, allowing the condensation of silicate and alumina grains. Radiation pressure on these dust grains can then drive a wind, especially when aided by prior pulsational levitation of gas \citep{1971ApJ...165..285G}. Dust-driven winds are supported by observations of IR excesses and extended envelopes in nearby RSGs such as Betelgeuse \citep{Dupree_2022} and VY~CMa \citep{DeBeck2016ALMAOO}. These winds often exhibit outflow velocities of 10–30~km\,s$^{-1}$ and mass loss rates between $10^{-6}$ and $10^{-4}$~\Ms~$\rm{yr}^{-1}$ \citep{2005A&A...438..273V}.

\subsection{Enhanced mass loss due to binary companion}
The majority of massive stars are believed to have experienced at least one episode of binary interaction in course of their evolution \citep{2012ARA&A..50..107L,2012Sci...337..444S}, although for RSGs, the binary fraction is estimated to be much lower both from observations \citep{2025MNRAS.539.1220D,2025ApJ...988...60D} and population synthesis predictions using the Binary Population and Spectral Synthesis code (BPASS;~\citealt{2017PASA...34...58E,2018MNRAS.479...75S,2020JOSS....5.1987S}). Irradiation effects due to heating or the ram pressure from the wind of a stellar companion or even an explosion, as well as tidal forces exerted by a binary companion (either a star, planet or a compact remnant) causing redistribution of the angular momentum inside the star, can trigger enhanced mass loss. 

Stars in binary systems may also accrete matter causing spin-up to critical rotation, thereby becoming partially unbound, resulting in mass loss through the inner Lagrangian point or even merging into a common envelope \citep{2017MNRAS.470.1788C,2021MNRAS.504L..51S}. Such evolutionary scenarios are also possible in X-ray binaries where the compact object (either a neutron star or black hole) accretes matter from its giant companion, making the latter lose mass through Roche lobe overflow, which could prevent it from evolving into a RSG and instead explode as a Type I b/c SN, a significant fraction of which are understood to originate from such envelope stripping by a binary companion \citep{2023MNRAS.523..221G,2024IAUS..361..465S}.


\section{RSG models with pulsations}\label{section: pulsating RSGs}

We model massive stars with initial masses between $12-20$~\Ms~ which evolve into red supergiants (RSGs) using the open-source stellar evolution code Modules for Experiments in Stellar Astrophysics \citep[MESA;][]{Paxton2011, Paxton2013, Paxton2015, Paxton2018, Paxton2019} - version 24.08.1. These models are computed hydrostatically from the pre-main sequence through to the end of core helium burning, and thereon hydrodynamically until core oxygen exhaustion  (for the full MESA implementation of our setup, refer to the Zenodo repository: \citealt{sengupta_2025_17605468}). We use a smaller nuclear network, \texttt{approx21\_cr60\_plus\_co56.net} \citep{1999ApJS..124..241T} until core carbon depletion, followed by a larger network, \texttt{mesa\_128.net} \citep{1999ApJS..124..241T,Timmes_2000} to better capture the energy generation in advanced phases of nuclear burning, due to limited computational resources. Our models also incorporate time-dependent convection as implemented in MESA by \cite{2023ApJS..265...15J} to better approximate the turbulent conditions within the convective envelopes of RSGs, than it is possible using the standard mixing length approach. We use MESA's inbuilt Dutch wind scheme until the end of core helium burning, scaled by a factor \texttt{dutch\_scaling\_factor}$= 0.2$, which gives RSG mass loss rates consistent with observations of large samples of RSGs in the Milky Way, the Magellanic Clouds \citep{2023A&A...676A..84Y,antoniadis_2024,2024AJ....167...51W} as well as M31, M33 \citep{2023ApJ...942...69M} and other nearby star clusters with coeval populations of RSGs from which mass-dependent mass loss rates have been inferred \citep{10.1093/mnras/stad1818,2024A&A...681A..17D}. Beyond core helium exhaustion, we switch to our custom mass loss prescription as described below in \ref{subsection : pdsw} to model the effect of radial pulsations on mass loss in RSGs ($\mathrm{T}_{\rm{eff}} < 4000~\mathrm{K}$). We also investigate late mass ejections in the subsequent evolution of our RSG models before CC, due to dynamical instabilities post-shock breakout in the pulsating stellar envelope, as described in \ref{subsection : deso}. In \ref{subsection : wind propagation} we describe the model for line-driven acceleration of the RSG wind generated through the various mass loss mechanisms mentioned above. Finally in \ref{subsection :  csm density} we calculate the CSM density profiles and the column density along line-of-sight assuming spherically symmetric wind using the accelerated velocity profile for comparison with few well-known Type II SNe (Section \ref{section : comparison}).

\subsection{Pulsation-Driven Superwind (PDSW)} \label{subsection : pdsw}

\begin{figure}[h]
    \centering
    \includegraphics[width=0.45\textwidth]{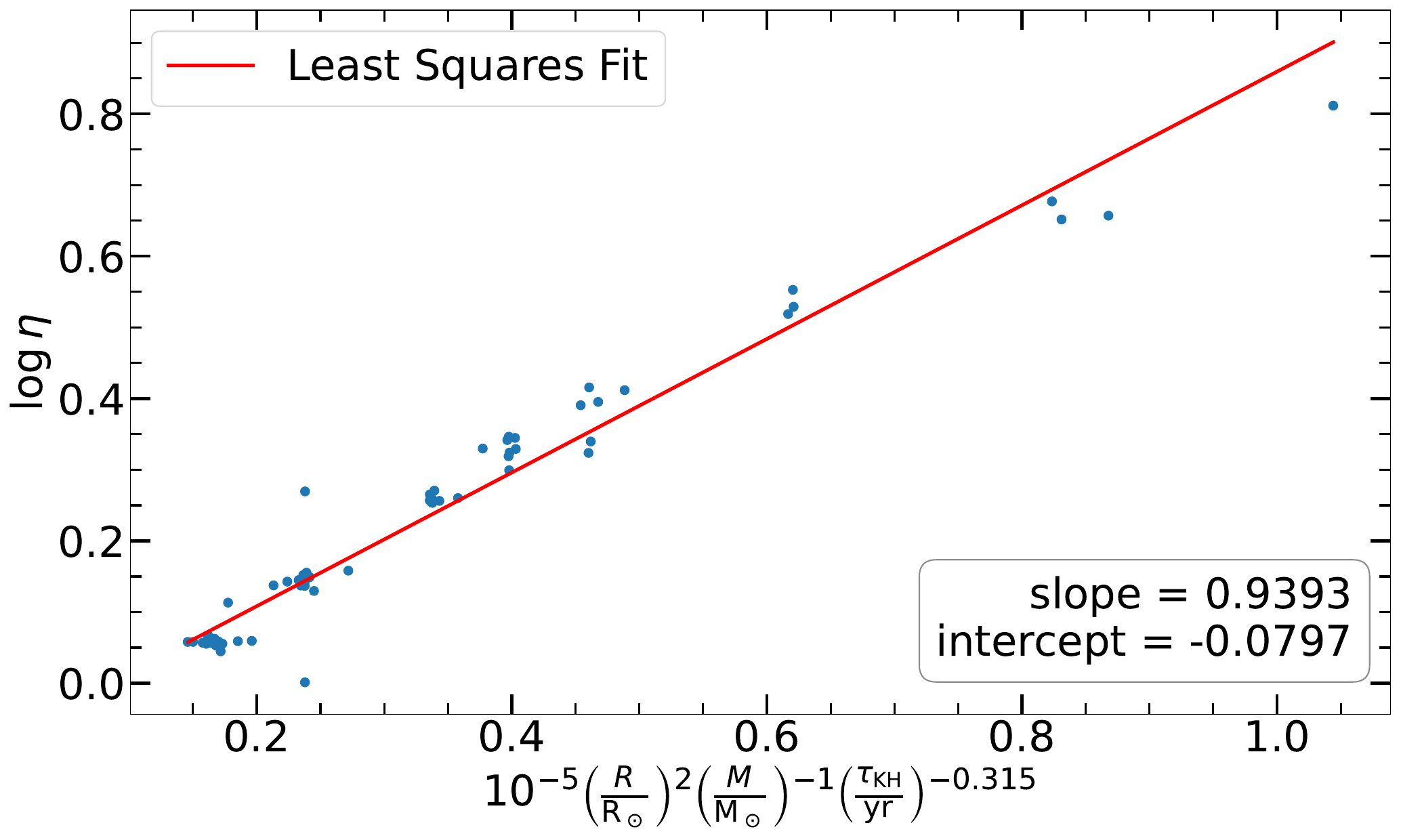}
    \caption{Growth rate of radial oscillations, $\eta$, as a function of stellar parameters (radius, mass, and Kelvin-Helmholtz time-scale of the envelope); the solid line represents a linear least-squares fit.}
    \label{fig:figure1}
\end{figure}

Red Supergiants (RSGs) often exhibit long-period variability \citep{2018ApJ...859...73S, 2019ApJS..241...35R, 2024arXiv240807874H} that has been associated with changes in stellar radii due to excitation of different modes of pulsational instabilities through the $\kappa$-mechanism \citep{2016IAUFM..29B.573S}. Theoretically, \cite{1997A&A...327..224H} demonstrated the growth of such pulsational instabilities in hydrodynamical evolutionary calculations with sufficient temporal resolution and found that the growth rates in their non-linear numerical calculations to be proportional to the luminosity ($\mathrm{L}$) to mass ($\mathrm{M}$) ratio. \cite{2010ApJ...717L..62Y} showed that the growth rate ($\eta$) of the amplitude of the surface velocities increases with $L/M$ ratio and/or decreasing thermal (Kelvin-Helmholtz) time scale of the envelope ($\tau_{KH, env}$), according to the relation: 

\begin{equation} \label{equation : 1}
\log \eta = C_1\left(\frac{R}{\mathrm{R}_\odot}\right)^2\left(\frac{M}{\mathrm{M}_\odot}\right)^{-1} \left(\frac{\tau_{\mathrm{KH,env}}}{\mathrm{yr}}\right)^{-0.315} - C_2,
\end{equation}

\noindent where $R$ is the stellar radius, $\tau_{KH, env} : = GMM_{env}/LR$ ($M_{env}$ is the mass of the H-rich envelope ) and $C_{1} = 9.219\times10^{-6},~C_{2} = 0.0393$ which are calibrated against their hydrodynamic models evolved until shock breakout at the surface. Figure~\ref{fig:figure1} shows a linear least-squares fit for the relation given by Equation~\ref{equation : 1} for hydrodynamic MESA model sequences with initial masses between $12 - 18$~\Ms~evolved with timesteps $\delta t\leq 0.01~\text{years}$ until shocks develop at the stellar surface, which gives $C_{1} = 9.393\times10^{-6}, C_{2} = 0.0797$, in close agreement with the values of \cite{2010ApJ...717L..62Y}.

\begin{figure}[h]
    \centering
    \begin{minipage}{0.49\textwidth}
        \centering
        \includegraphics[width=\textwidth]{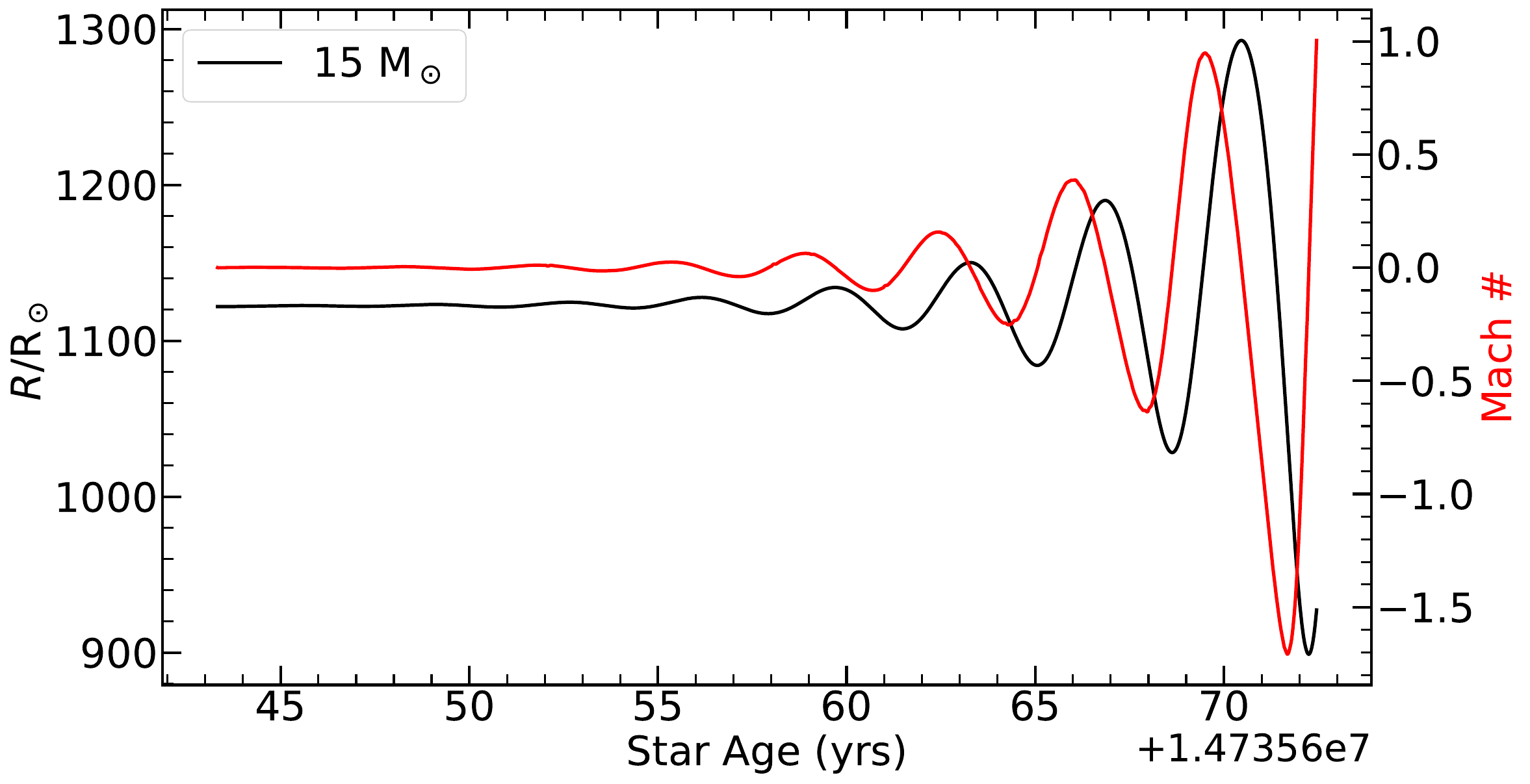}
    \end{minipage}
    \hfill
    \begin{minipage}{0.45\textwidth}
        \centering
        \includegraphics[width=\textwidth]{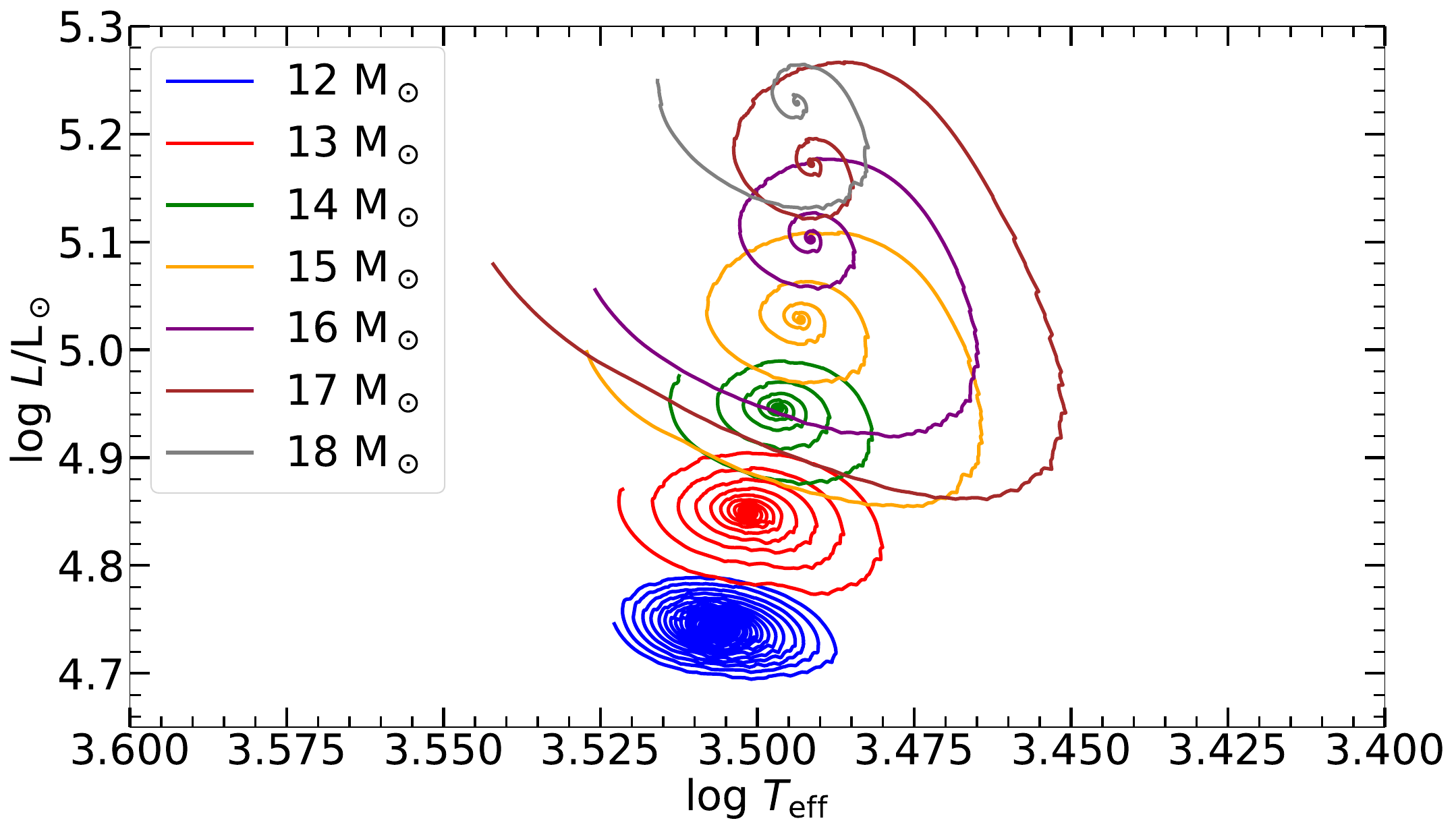}
    \end{minipage}
    \caption{(Top) Evolution of the stellar radius (black curve plotted on the left axis) in a set of hydrodynamic MESA model sequences with initial mass $M_\mathrm{init} = 15~\mathrm{M}_\odot$ in the RSG phase until shock breakout at the surface when Mach Number  ($ := v_{surf}/c_{sound}$) exceeds unity (shown in red on the right axis). (Bottom) Corresponding evolution of a set of MESA models with varying initial masses between $12 - 18\ M_\odot$ in the HR diagram.}
    \label{fig:figure2}
\end{figure}

 The top panel of Figure~\ref{fig:figure2} shows a typical hydrodynamical model for a MESA run with initial mass $M_\mathrm{init} = 15$~\Ms~illustrating the growth of the radial pulsations until shock break-out at the surface where the Mach number reaches unity. The bottom panel of Figure~\ref{fig:figure2} shows the corresponding evolution of a set of MESA runs with varying initial masses between $12 - 18$~\Ms~in the Hertzspring-Russell (HR) diagram. Capturing these radial pulsations with the typical time period, $P\sim 2-5~yrs$, requires using much smaller time steps \citep{Paxton2013}, that makes modeling such pulsations a challenging task for evolutionary codes which typically use much larger time steps during the onset and growth of these pulsations \citep{Paxton2013,2025A&A...703A..61B}. We do not attempt to resolve these pulsations for our full evolutionary runs (from pre main-sequence until core neon/oxygen depletion) for computational feasibility but incorporate potential impact of the growth of such pulsations observed in advanced phases (beyond core-helium burning) on the mass loss history of our stellar models in a phenomenological approach, similar to that of \cite{2010ApJ...717L..62Y}. To do this, we evolve a set of hydrostatic MESA models until end of core-helium burning using the standard Dutch wind scheme in MESA, scaled down by a factor of $0.2$ for reasons explained in Section~\ref{section: pulsating RSGs}. Thereafter, each run is continued with the hydrodynamic term on and replacing the wind mass loss rates with the "superwind" (SW) mass loss prescription of \cite{2010ApJ...717L..62Y} given by  
 \begin{equation} \label{equation : 2}
\dot{M} = \eta^{\alpha} \dot{M}_\mathrm{JNH88},
\end{equation}
for $\eta>1$ and $\mathbf{T_{\rm{eff}}<4000~\rm{K}}$, using the growth rate expression as given in Equation \ref{equation : 1} with our calibrated constants. Here, $\dot{M}_\mathrm{JNH88}$ is the mass loss rate from de \cite{1988A&AS...72..259D} scaled down by a factor of $0.2$, and  $\alpha$ is a free parameter which sets the strength and consequently the duration of the SW-phase (Section~\ref{subsection : deso}). Unlike \cite{2010ApJ...717L..62Y} who used a fixed $\alpha~=~1.75$, we conduct a thorough exploration of the dependence of the SW on both the initial mass $\mathrm{M}_{init}=12-20$~\Ms~and value of $\alpha=0-5$, motivated by observational evidence for enhanced mass loss in late stage of evolution of RSG progenitors of Type II SNe.
\begin{figure}[h]
    \centering
    \includegraphics[width=0.45\textwidth]{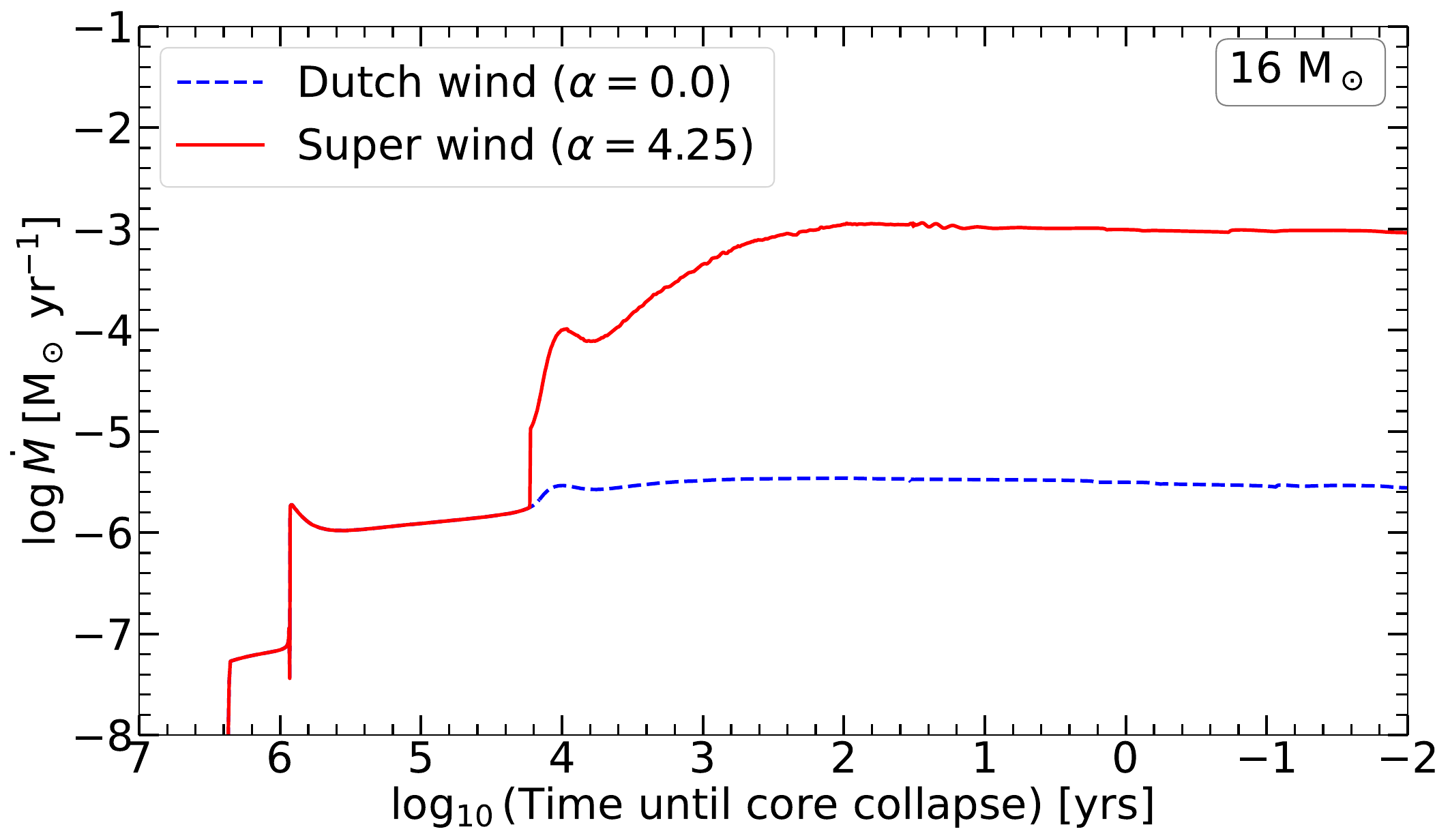}
    \caption{Mass loss histories of two model sequences with $M_{\rm{init}}=16~\rm{M}_\odot$ for $\alpha=0.0$ (dashed blue line) which corresponds to the Dutch wind in MESA, scaled down by a factor of 0.2 (see Section~\ref{section: pulsating RSGs}), and $\alpha=4.25$ (solid red line) which gives an enhanced "superwind" with $\dot M\sim10^{-3}$~\Ms~$\rm{yr}^{-1}$ over the last $\sim10^{3}~\rm{yrs}$.}
    \label{fig:figure3}
\end{figure}

Figure \ref{fig:figure3} illustrates a sequence of such MESA models for $M_\mathrm{init} = 16$~\Ms,~$\alpha=4.25$ showing the evolution of the mass loss rate as a function of the logarithm of the time to CC, measured from right to left. Thus, the zero of the x-axis represents one year from CC, approximately corresponding to the duration of oxygen-burning in our models, which is the final phase up to which we have evolved all our model sequences. It is readily seen that the model with $\alpha=4.25$ evolves with an enhanced mass loss rate over the last $\sim10^3$ years, which is approximately two orders of magnitude higher compared to the model sequences evolved with the Dutch prescription in MESA, scaled down by a factor of $0.2$, obtained by setting $\alpha=0$ in Equation \ref{equation : 2}. At lower values of either the initial mass or $\alpha$, mass loss is less enhanced compared to the corresponding Dutch wind values but the qualitative picture is similar for lower mass progenitors with similar or lower values of $\alpha$ (see Figure \ref{fig:5}).

\subsection{Dynamical ejections post shock break-out} \label{subsection : deso}
The SW prescription of \cite{2010ApJ...717L..62Y} is strictly applicable only to the exponential growth phase of radial pulsations, which subsequently get damped by shocks that develop within the RSG envelope where the radial velocity of the loosely bound layers becomes supersonic. \cite{clayton2018a} investigated the problem of dynamical stability of RSG envelopes in the shock-dominated regime, in terms of the pressure-weighted, volume averaged first adiabatic exponent \citep[see Section 3.4 in~][]{2017MNRAS.470.1788C} whose value less than $\mathbf{4/3}$ at any point within the stellar envelope is indicative of instability to dynamical ejections from that point outwards \citep{2001ApJ...558..780L} - commonly occurring in regions above the hydrogen and helium ionization zones. Detailed shock-resolving hydrodynamic MESA models of \cite{clayton2018a} showed launch of ejections upon the surface breakout of compression shocks followed by sufficiently fast expansion that raises material on to an escape trajectory. \cite{clayton2018a} implemented a wind-mass loss scheme in order to excise any such contiguous unbound sub-surface layer of mass $m$  (with radial velocity exceeding the local escape velocity), by applying a mass loss rate of $100m$~$\rm{yr}^{-1}$ corresponding to a time-scale which is at least one order of magnitude below both the dynamical and thermal time-scales of the star. A subset of their extensive suite of hydrodynamic simulations of such shock-driven mass loss resulted in episodes of repeated mass ejection events lasting for upto $100$-s of years. Almost all cases that converged upon a solution after showing such episodes of repeated mass ejections had $\log((\mathrm{L}/\mathrm{L}_\odot)/(\mathrm{M}/\mathrm{M}_\odot))\gtrsim4.1$, for which \cite{clayton2018a} extracted time-averaged mass loss rates to propose the following mass loss prescription:
 \begin{equation} \label{equation : 3}
\log(\dot{M}/\mathrm{M_\odot}~\rm{yr}^{-1}) = 5.93\times \log[(L/\mathrm{L}_\odot)/(M/\mathrm{M}_\odot)]-26.6,
\end{equation}
above a cut-off value of $\log((L/\mathrm{L}_\odot)/(M/\mathrm{M}_\odot))$ between $4.1-4.15$. Hence, we employ this mass loss recipe to model the possibility of such repeating mass ejections occurring post shock break-out in pulsationally unstable RSG envelopes when the luminosity to mass ratio of the star exceeds the cut-off value of $\log((L/\mathrm{L}_\odot)/(M/\mathrm{M}_\odot))=4.15$, which is on the conservative side, given the uncertainties in determining the exact value of this cut-off \citep{clayton2018a}. We do not consider any other mass loss mechanism in this regime, so as to avoid duplicating effects of any similar mechanism(s) triggering mass loss in the final years leading to CC. Furthermore, the above recipe does not account for other dynamically unstable scenarios which lead to single ejection episodes, also identified in the simulations of \cite{clayton2018a}, that occur on a much shorter timescale to be accounted for in our evolutionary calculations.

\begin{figure}[h]
    \centering
    \includegraphics[width=0.45\textwidth]{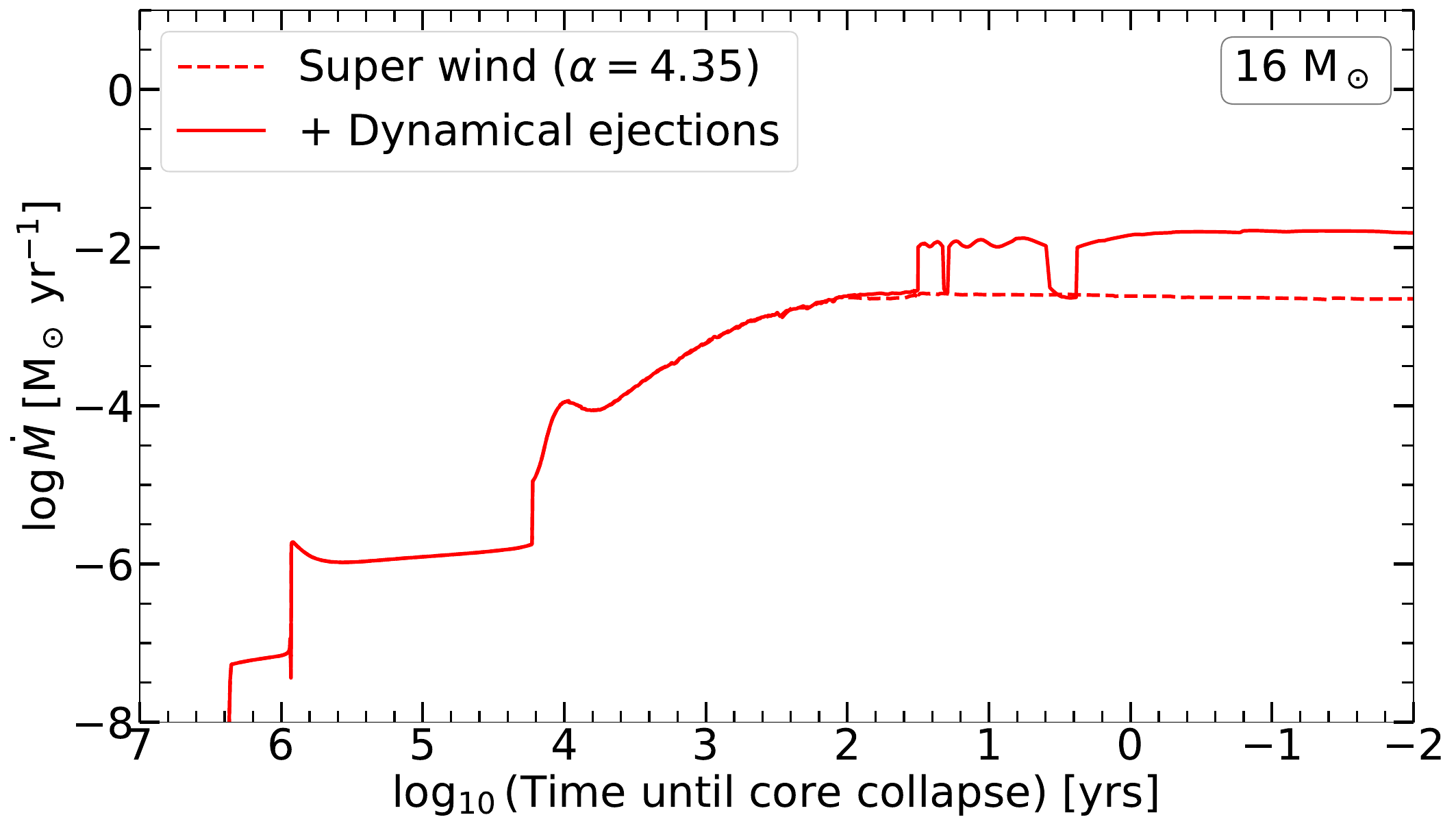}
    \caption{Mass-loss histories of two model sequences as described in Section \ref{subsection : pdsw} with $M_{\rm{init}}=16~\rm{M}_\odot$ for $\alpha=4.35$ (in dotted red) and including dynamical mass ejections (in solid red) which gives episodic enhancements in mass-loss with $\dot M\sim10^{-2}$~\Ms~$\rm{yr}^{-1}$ over the last $\sim30~\rm{yrs}$ prior to CC.}
    \label{fig:figure4}
\end{figure}

Figure \ref{fig:figure4} shows such a model sequence of initial mass $M_\mathrm{init} = 16$~\Ms~evolved with both the prescriptions given by Equations \ref{equation : 2} (for $\alpha=4.35$) and \ref{equation : 3} implemented in MESA, with a transition at $\log((L/\mathrm{L}_\odot)/(M/\mathrm{M}_\odot))=4.15$. It can be seen that the onset of the dynamical ejections marks a significant enhancement in the mass loss rates (around an order of magnitude) compared to the superwind phase, which has important implications for the CSM formed by these winds close to the star, as in this model the ejections occur less than $100$~years from CC.

\begin{figure*}[ht!]
    \centering
    \begin{subfigure}[t]{0.48\textwidth}
        \includegraphics[width=\textwidth]{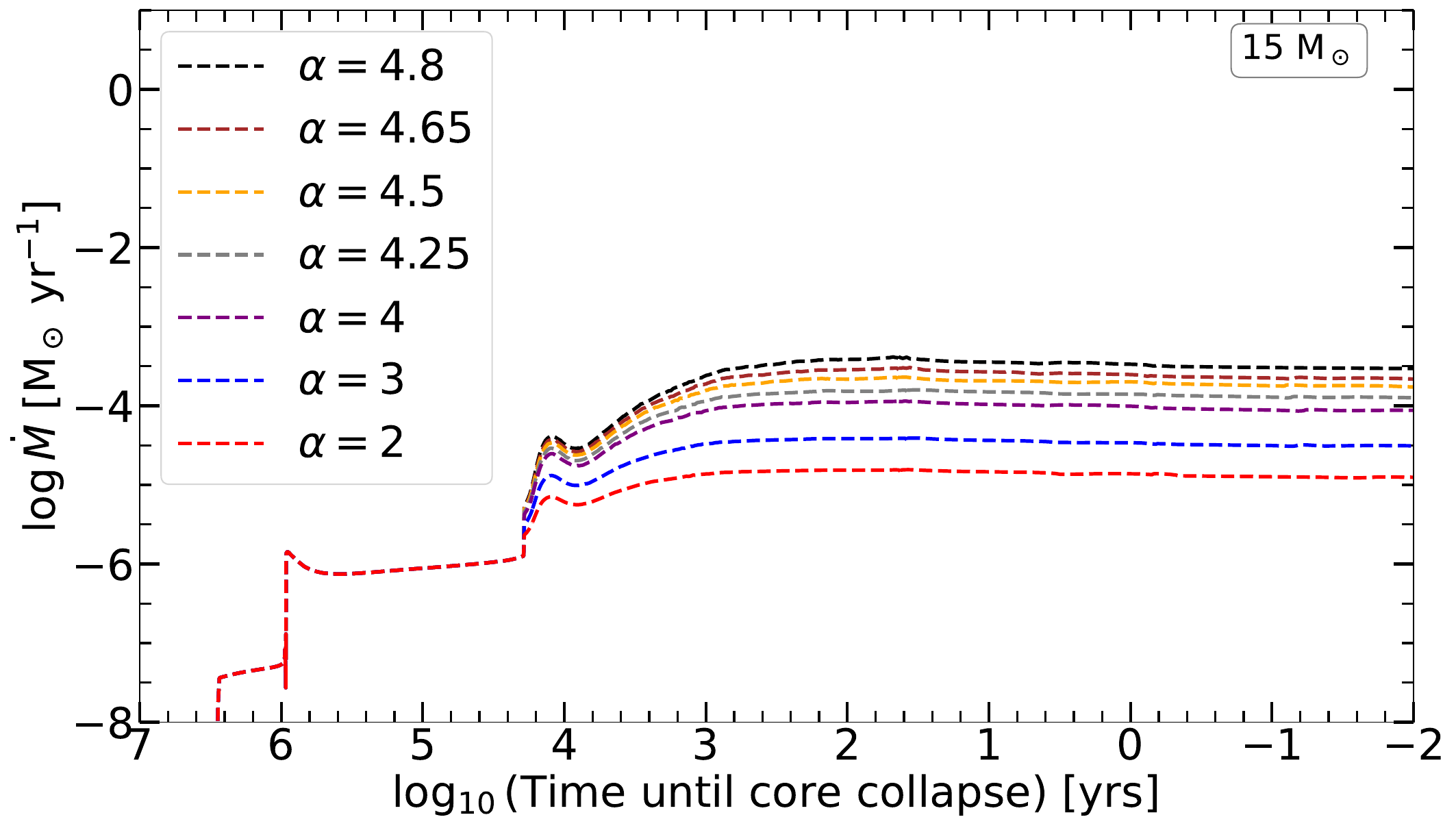}
        \label{fig:5a}
    \end{subfigure}
    \hfill
    \begin{subfigure}[t]{0.48\textwidth}
        \includegraphics[width=\textwidth]{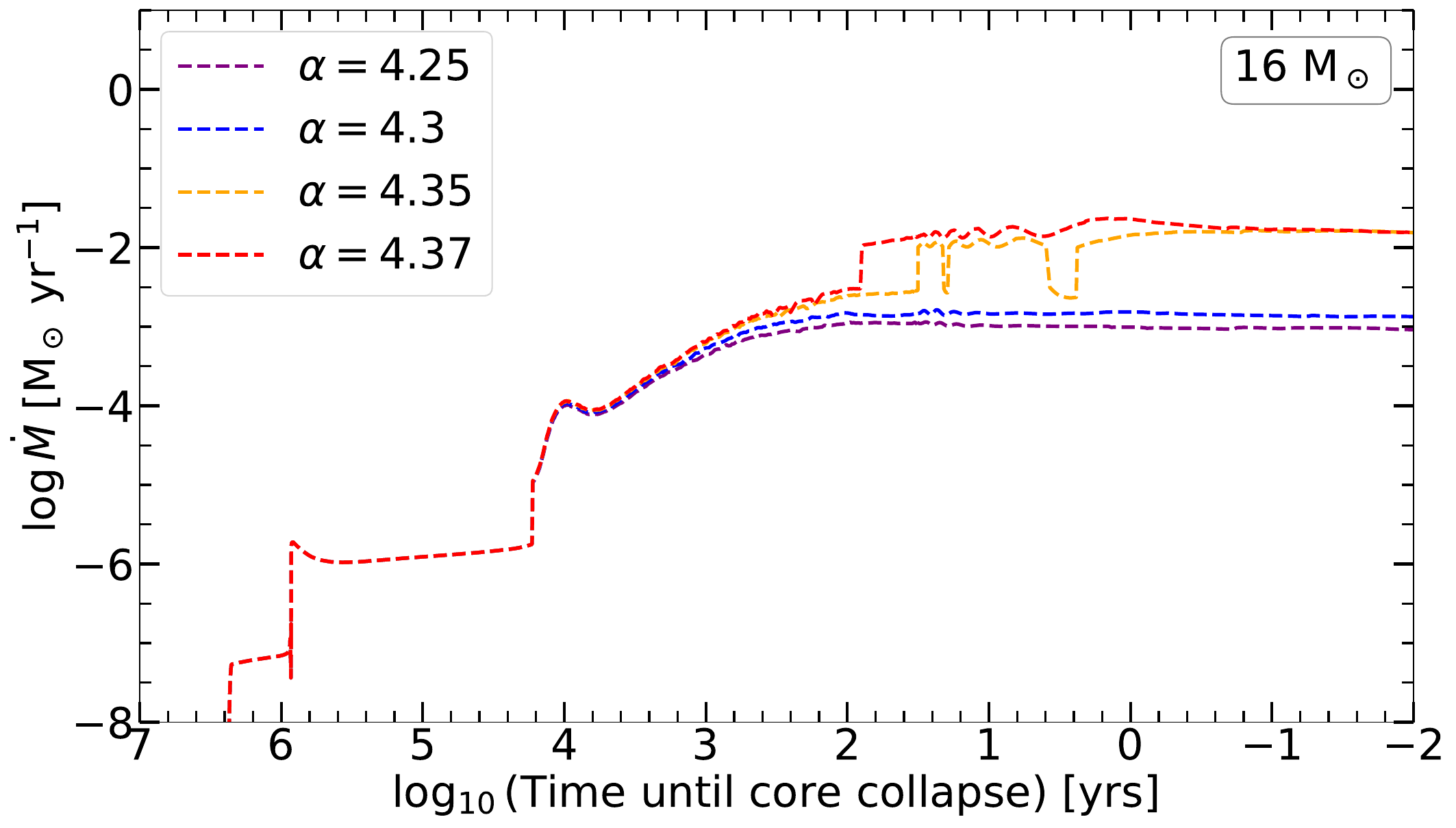}
        \label{fig:5b}
    \end{subfigure}
    
    \vspace{1em}
    
    \begin{subfigure}[t]{0.48\textwidth}
        \includegraphics[width=\textwidth]{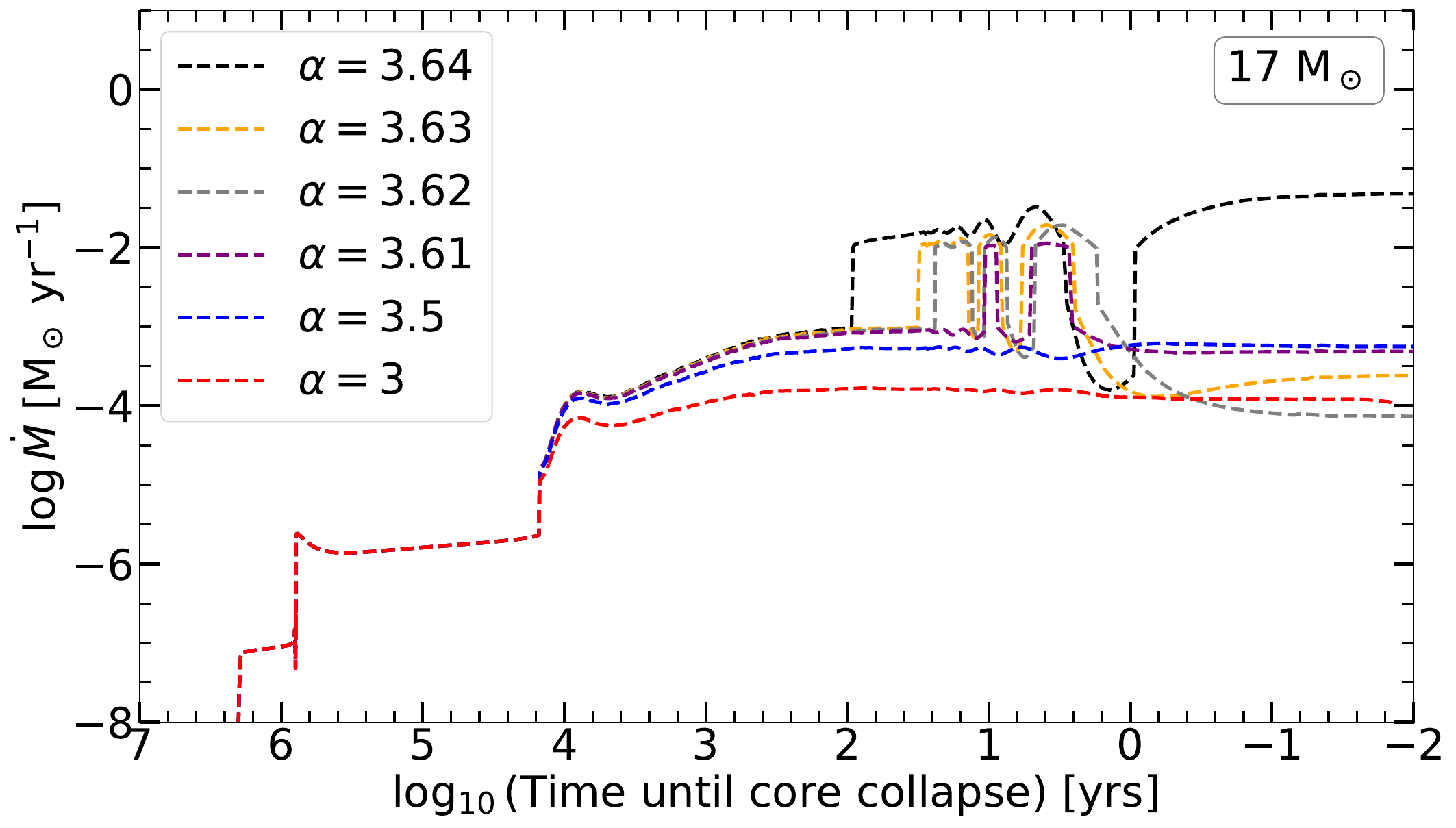}
        \label{fig:5c}
    \end{subfigure}
    \hfill
    \begin{subfigure}[t]{0.48\textwidth}
        \includegraphics[width=\textwidth]{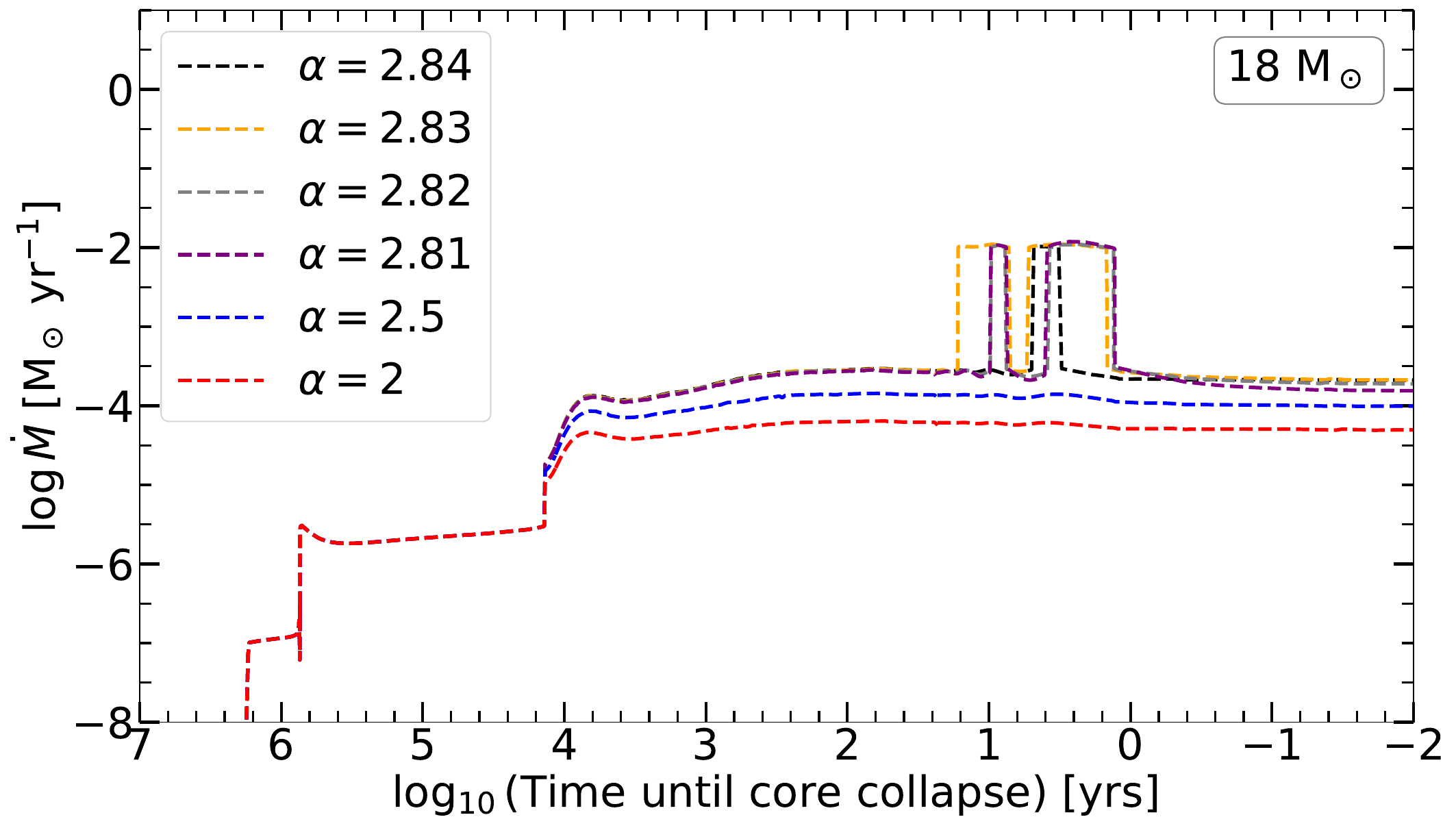}
        \label{fig:5d}
    \end{subfigure}
    
    \caption{
        Evolution of MESA model sequences with varying initial masses showing the mass-loss history due to PDSWs and dynamical ejections as described in Sections \ref{subsection : pdsw} and \ref{subsection : deso} respectively: \textit{top left panel} shows the mass loss history of a sequence of MESA models with $M_{\rm{init}}=15$~\Ms~ prior to CC, showing enhanced mass-loss rates of $\dot M\gtrsim10^{-5}$~\Ms~$\rm{yr}^{-1}$ for $\alpha\ge2$, due to PDSWs. These models do not reach the regime of dynamical ejections; \textit{top right panel} shows MESA models with $M_{\rm{init}}=16$~\Ms~exhibiting episodic mass loss due to dynamical ejections for $\alpha\gtrsim4.35$ following a superwind phase, with average mass loss rates exceeding $10^{-2}$~\Ms~$\rm{yr}^{-1}$ within $100$~yrs of CC; \textit{bottom left panel} shows MESA model sequences for $M_{\rm{init}}=17$~\Ms~, showing similar episodes of elevated mass loss due to dynamical ejections occur at $\alpha\gtrsim3.6$ with earlier onset of such phases observed at higher values of $\alpha$; \textit{bottom right panel} shows MESA models for $M_{\rm{init}}=18$~\Ms~, showing onset of the dynamical ejections occuring at $\alpha\gtrsim2.8$ between $1-20$~yrs of CC.
    }
    \label{fig:5}
\end{figure*}

Figure~\ref{fig:5} shows a set of evolutionary calculations using both the mass loss prescriptions for PDSW (Equation~\ref{equation : 2}) and subsequent dynamical ejections (Equation~\ref{equation : 3}). From our grid of MESA models, we find that only model sequences with initial mass $M_{\rm{init}}>15$~\Ms~ evolve to enter the regime for which dynamical ejections are expected (as described in \ref{subsection : deso}) for $\mathbf{\alpha=2-5}$. Furthermore, depending on the initial mass, there is a cut-off value of $\alpha$ above which the dynamical ejections lead to elevated mass loss ($\dot M\gtrsim10^{-2}$~\Ms~$\rm{yr}^{-1}$). 

\subsection{Wind propagation: CSM density, integrated column density} \label{subsection : wind propagation}
In this section, we describe our approach to construct the CSM surrounding the RSG progenitor modeled using MESA as described thus far. To do so, we need a model for the wind propagating outward from the stellar surface which we describe below (\ref{subsubsection : wind propagation}) along with the numerical procedure used to construct the CSM density (\ref{subsection :  csm density}) as well as the column density along the line of sight (\ref{subsection :  col density}).

\subsubsection{Acceleration of the RSG wind} \label{subsubsection : wind propagation}
Following \cite{2018MNRAS.476.2840M}, we consider acceleration of the stellar wind -- which starts with a velocity $v_o$ at the launch radius $r=R_*$, which is is assumed to be just the stellar radius -- out to large radii where the wind velocity gradually reaches a constant terminal speed, $v_\infty$. Since the exact mechanism of acceleration of the wind remains unknown, the kinematics is usually parametrized in form of the following simple $\beta$-law for the wind velocity \citep{1999isw..book.....L}: 

\begin{equation} \label{equation : 4}
v_w(r) = v_0 + (v_\infty-v_0)\left(1-\frac{R_*}{r}\right)^\beta,
\end{equation}
where the value of $\beta$ determines the shape of the wind profile. Usually, both $v_\infty$ and $\beta$ are obtained through spectral line fitting for which $0.7\lesssim\beta\lesssim4$ has been used in the literature \citep{2000ARA&A..38..613K}. RSGs known to experience slower wind acceleration as compared to hotter (OB) stars i.e. $\mathbf{\beta\gtrsim1}$ \citep{1996ApJ...466..979B,2010ASPC..425..181B}. For the following comparisons of our results with CSM properties of observed SNe (Section \ref{section : comparison}), we adopt a constant value of $\beta=1.2$, consistent with hydrodynamical simulations \citep{galaxies11030068} as well as observations of single RSGs (\citealt{2010ASPC..425..181B}; see Figure~\ref{fig:figure11a} in Appendix \ref{beta-law}) while the values of the terminal velocity, $v_\infty$, are used specific to each SN  as inferred from fitting of spectral lines.

\subsubsection{CSM density calculation} \label{subsection :  csm density}
Using the accelerated wind profile given by Equation \ref{equation : 4}, we propagate the wind outward by integrating along the radial direction using a fourth-order Runge-Kutta scheme, in order to compute the CSM density, $\rho(r)$ at a radial distance $r$ from the stellar surface. Assuming spherical symmetry, the CSM density is given by
\begin{equation} \label{equation : 5}
\rho(r) = \frac{\dot M(t)}{4\pi r^2v_w(r)},
\end{equation}
where $\dot M(t)$ is the mass loss rate of the the progenitor computed at a time $t$ before CC, during which the wind has propagated out to a radius $r$ from the surface of the star. The launch speed $v_0$ of the wind at the stellar surface is set by requiring the CSM density to be connected smoothly at the stellar surface.

\subsubsection{Column density along line of sight calculation} \label{subsection :  col density}
The column density of hydrogen along the line of sight is computed using the CSM density obtained as described in \ref{subsection :  csm density} as
\begin{equation} \label{equation : 6}
N_H(r) = \int_r^\infty n(r')dr',
\end{equation}
where the number density of hydrogen at a radial distance $r$ within the CSM is given by 
\begin{equation} \label{equation : 7}
n(r) = \frac{\rho(r)}{\mu m_p},
\end{equation}
with the mean molecular weight $\mu\simeq1.3$ for hydrogen, and the proton mass $m_p=1.6726\times10^{-24}~gm$. The computed values of the column densities are compared in \ref{section : comparison} with available measured values for well-studied SNe (1998S, 2005ip, 2020ywx, 2023ixf), using observed estimates of terminal wind speeds, $v_\infty$, required to compute the CSM densities, as per Equation~\eqref{equation : 5}.

\section{Comparison with observed Type II SN\MakeLowercase{e}} \label{section : comparison}
In this section, we construct the CSM around RSG progenitors of CCSNe of varying initial masses as described in \ref{subsection :  csm density} and \ref{subsection :  col density}, and compare our results with well-studied Type II supernovae with observed estimates of mass loss rates and/or measured values of column densities. Our goal is to ascertain the ability of our models to reproduce late phase enhanced mass loss pre-explosion that may create dense CSM around the SNe as inferred from post-explosion observations for the selected events presented below.

\subsection{SN 2023ixf}

\begin{figure}[H]
    \begin{minipage}[t]{0.49\textwidth}
        \raggedleft
        \begin{subfigure}[t]{0.99\textwidth}
            \includegraphics[width=\textwidth]{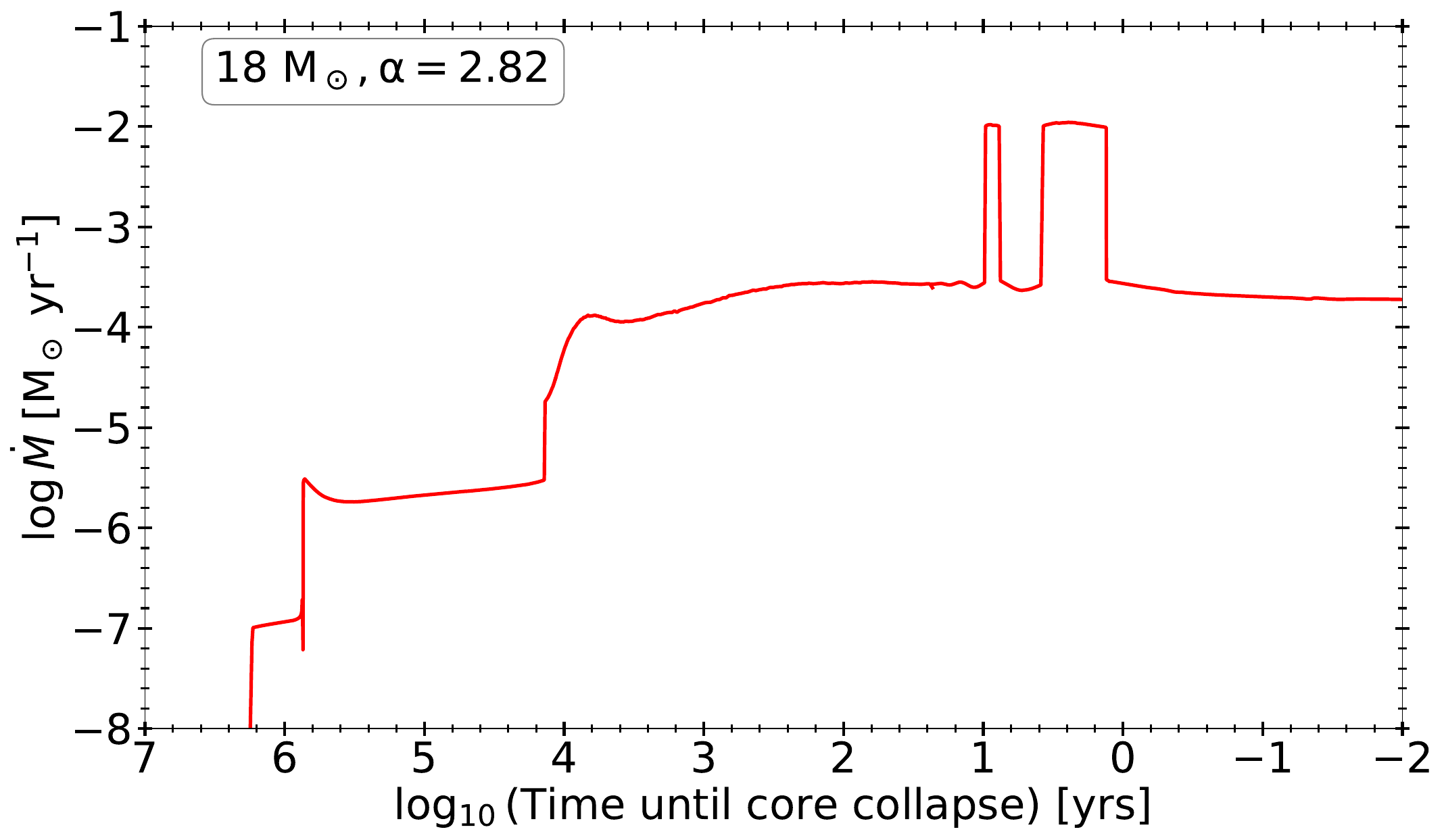}
            \label{fig:10a}
        \end{subfigure}
        \vfill
        \begin{subfigure}[t]{0.99\textwidth}
            \includegraphics[width=\textwidth]{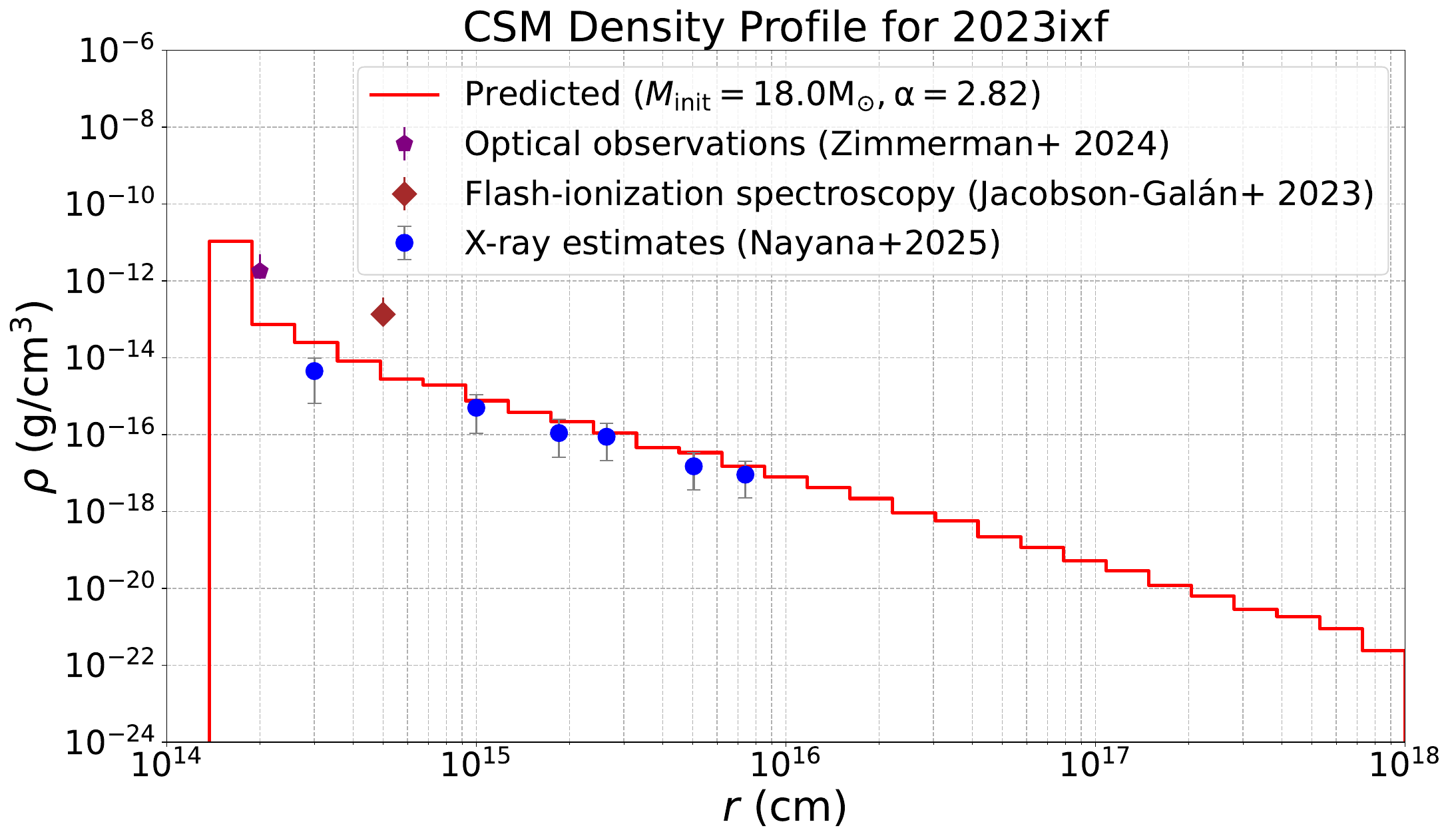}
            \label{fig:10b}
        \end{subfigure}
        \vfill
        \begin{subfigure}[t]{0.99\textwidth}
            \centering
            \includegraphics[width=\textwidth]{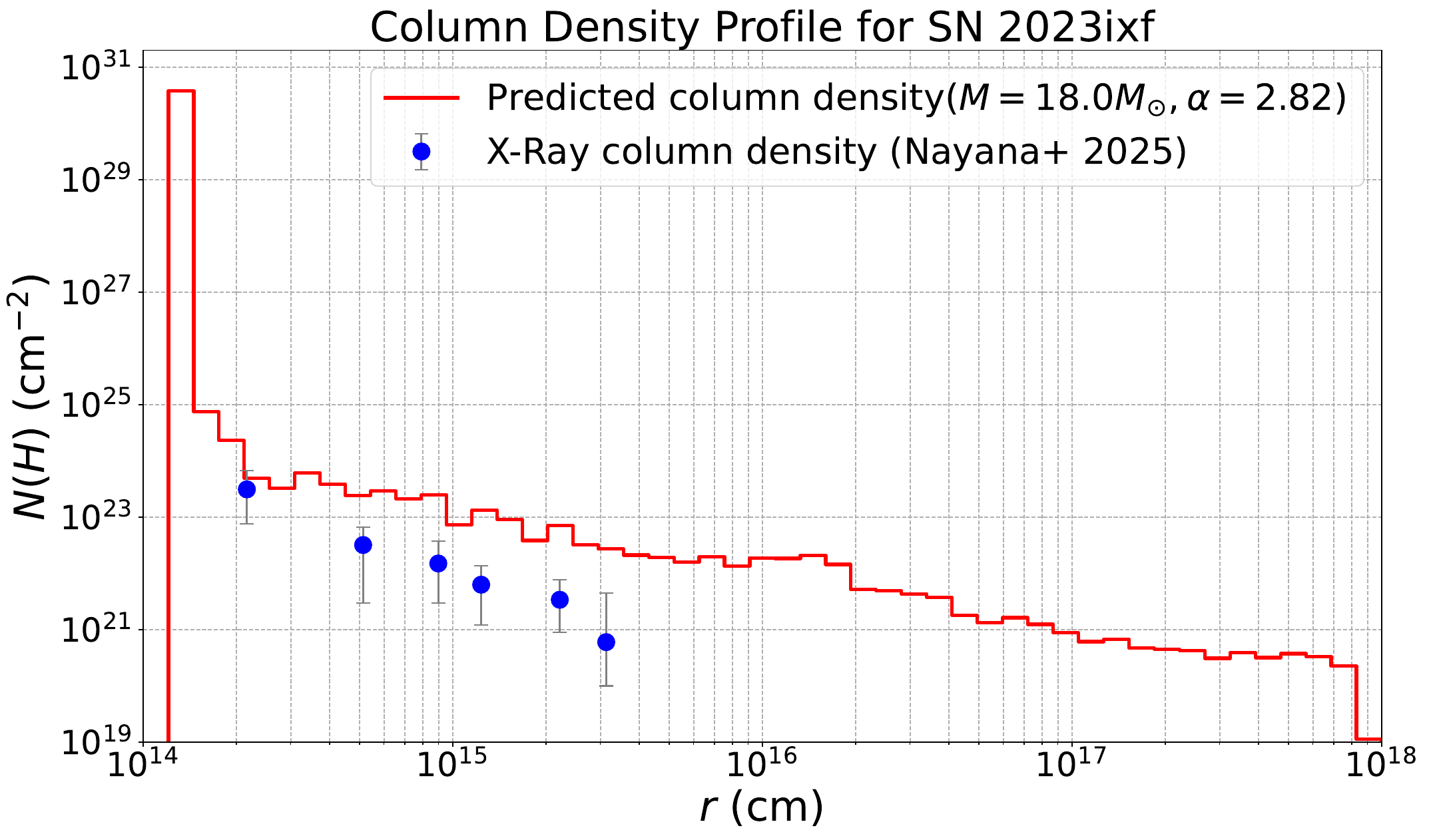}
            \label{fig:10c}
        \end{subfigure}

        \caption{
            Evolution of mass-loss rates of our $M_{\rm{init}}=18$~\Ms,~$\alpha=2.82$ MESA model sequences along with corresponding CSM densities compared with observationally inferred values for SN 2023ixf as well as the column density of neutral hydrogen compared to observed estimates \citep{2025ApJ...985...51A}: 
            \textit{top panel} shows the mass-loss history of the $ M_{\rm{init}}=18$~\Ms, $\alpha=2.82$ model, showing a superwind phase starting $\sim10^{4}$ years before core collapse and followed by the late onset dynamical ejections in the final decade with mass loss rates up to $10^{-2}$~\Ms~$\rm{yr}^{-1}$;
            \textit{middle panel} shows the CSM density profile for the model sequence shown in the top panel (solid red curve), compared with inferred densities from multi-wavelength observations for SN~2023ixf \citep{2023ApJ...954L..42J,2024Natur.627..759Z,2025ApJ...985...51A};
            \textit{bottom panel} shows the column density of neutral hydrogen (solid red) calculated using the model CSM profile shown in the middle panel, compared to the measured column densities (in solid blue) from X-ray observations \citep{2025ApJ...985...51A}.
        }
        \label{fig:10}
    \end{minipage}
\end{figure}

SN~2023ixf is one of the nearest Type~II SNe observed in recent years, with extensive follow-up providing stringent constraints on the structure and density of its CSM \citep{2025arXiv250708078J}. Flash-ionization signatures in its spectra indicate that its red supergiant (RSG) progenitor underwent enhanced mass loss ($\dot M\sim10^{-2}$~\Ms~$\rm{yr}^{-1}$) during the final decade before CC, producing a compact dense shell (CDS) around it \citep{2023ApJ...952L..30J, 2023ApJ...954L..42J, 2024Natur.627..759Z, Smith2023ixf, 2024ApJ...973L..47B}. Multi-epoch NuSTAR, XMM-Newton, Swift-XRT and Chandra X-ray observations \citep{Chandra_2024} as well as radio follow-up with the NOEM, VLA, GMRT \citep{2025ApJ...985...51A} infer mass loss rates of $\sim10^{-4}$~\Ms~$\rm{yr}^{-1}$ at around $(0.4-1.5)\times10^{15}$~cm, indicating a variable mass loss rates over the last $\sim200$~yrs of the progenitor's life. 
To model this scenario, in the top panel of Figure~\ref{fig:10}, we show our \texttt{MESA} model sequence for a progenitor of initial mass $M_\mathrm{init} = 18$~\Ms~with a superwind parameter $\alpha = 2.82$. This model enters the regime of dynamical ejections $~11$~yrs prior to CC. The top panel of Figure~\ref{fig:10} shows the mass loss history of this model sequence, which shows the superwind phase starting at $10^{4}$~yrs before CC, with $\dot{M} \gtrsim 10^{-4}$~\Ms~$\rm{yr}^{-1}$, followed by episodes of dynamical ejections with averaged mass loss rates exceeding $10^{-2}$~\Ms~$\rm{yr}^{-1}$ in the final decade before explosion.

Using a terminal wind speed of $v_\infty=20\pm5~\rm{km~s}^{-1}$ measured from He I emission \citep{2025ApJ...984...71D,2025arXiv250811747J}, we compare our model predictions with inferred CSM densities from multi-epoch observations across optical, X-ray and radio wavelengths \citep{2025ApJ...985...51A}, as shown in the middle panel of Figure~\ref{fig:10}. We obtain good agreement with both the early and late epoch observations, using the mass loss history of our $18$~\Ms,~$\alpha=2.82$ model sequence that leads to high CSM densities ($\gtrsim10^{15}~\mathrm{g~cm}^{-3}$) close to the progenitor surface ($r\leq10^{15}$~cm).

In the bottom panel of Figure~\ref{fig:10}, we compare the column density of neutral hydrogen, computed using Equation~\eqref{equation : 7}, with measured X-ray values as reported in \cite{Chandra_2024} and \cite{2025ApJ...985...51A}, for which we use the evolution of the shock radius with time as per the estimates of \cite{Chandra_2024}, to convert the epochs into $r$. We observe that, in general, we overpredict the column densities at most epochs where we have observed X-ray measurements. This could arise due to possible global asymmetries in the CSM \citep{2025A&A...694A.319K}, as hinted by the rapid spectropolarimetric properties at early times in the ejecta \citep{2023ApJ...955L..37V,2025ApJ...982L..32S} and/or from clumps of material along the line of sight, as indicated by the detection of Fe emission in the X-ray spectra \citep{2025ApJ...985...51A}. While such effects have been attributed to the ejecta geometry \citep{2024A&A...687L..20F} and/or binary interaction \citep{10.1093/mnras/staf443}, regardless of the exact cause of their origins, either scenario would result in underestimation  of the integrated column densities along the line of sight in observations, which would make them consistent with our systematically higher model estimates.

\subsection{SN 2020ywx} \label{subsection: 2020ywx}
\begin{figure}[H]
    \begin{minipage}[t]{0.49\textwidth}
        \raggedleft
        \begin{subfigure}[t]{0.99\textwidth}
            \includegraphics[width=\textwidth]{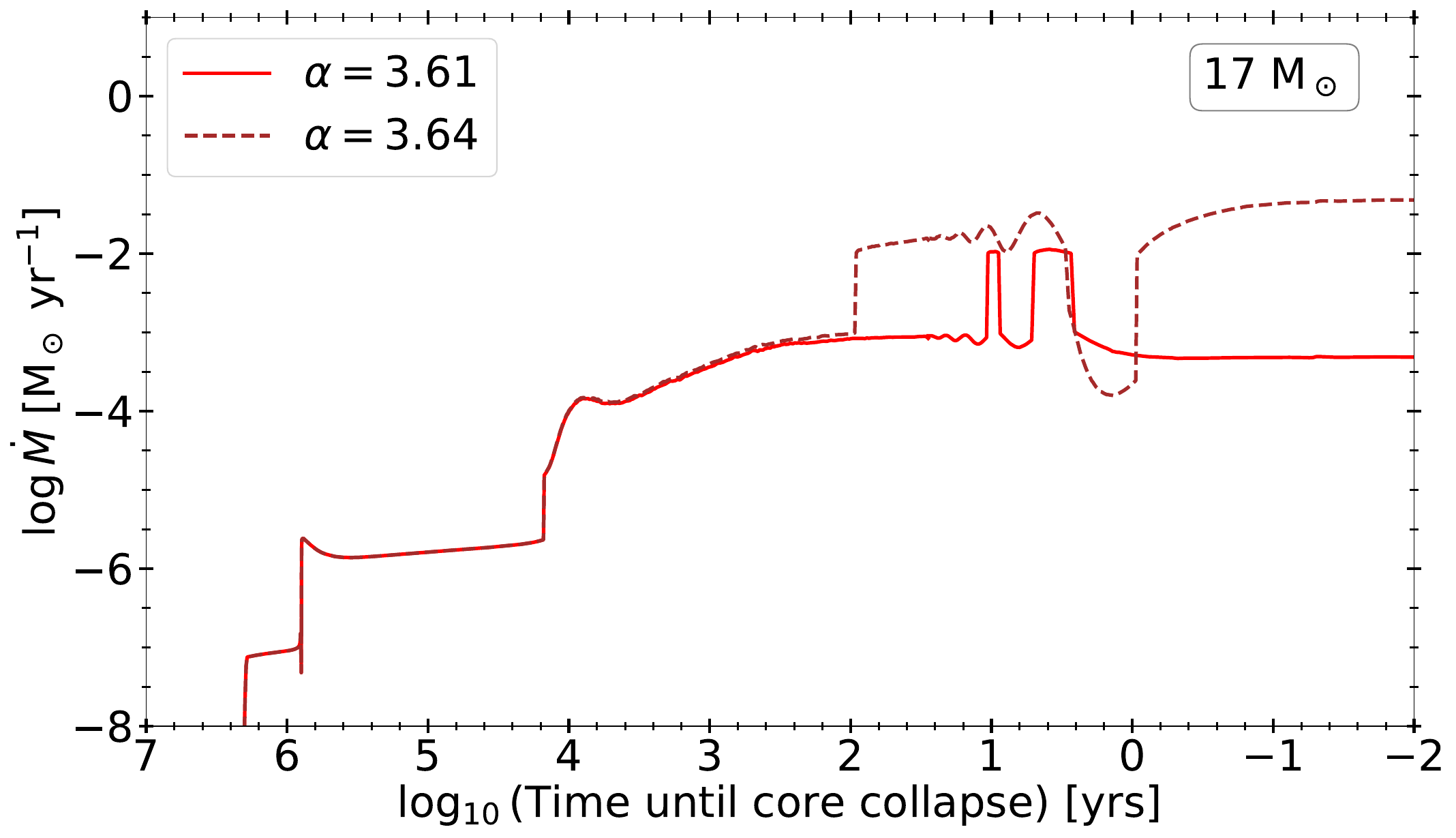}
            \label{fig:9a}
        \end{subfigure}
        \vfill
        \begin{subfigure}[t]{0.99\textwidth}
            \includegraphics[width=\textwidth]{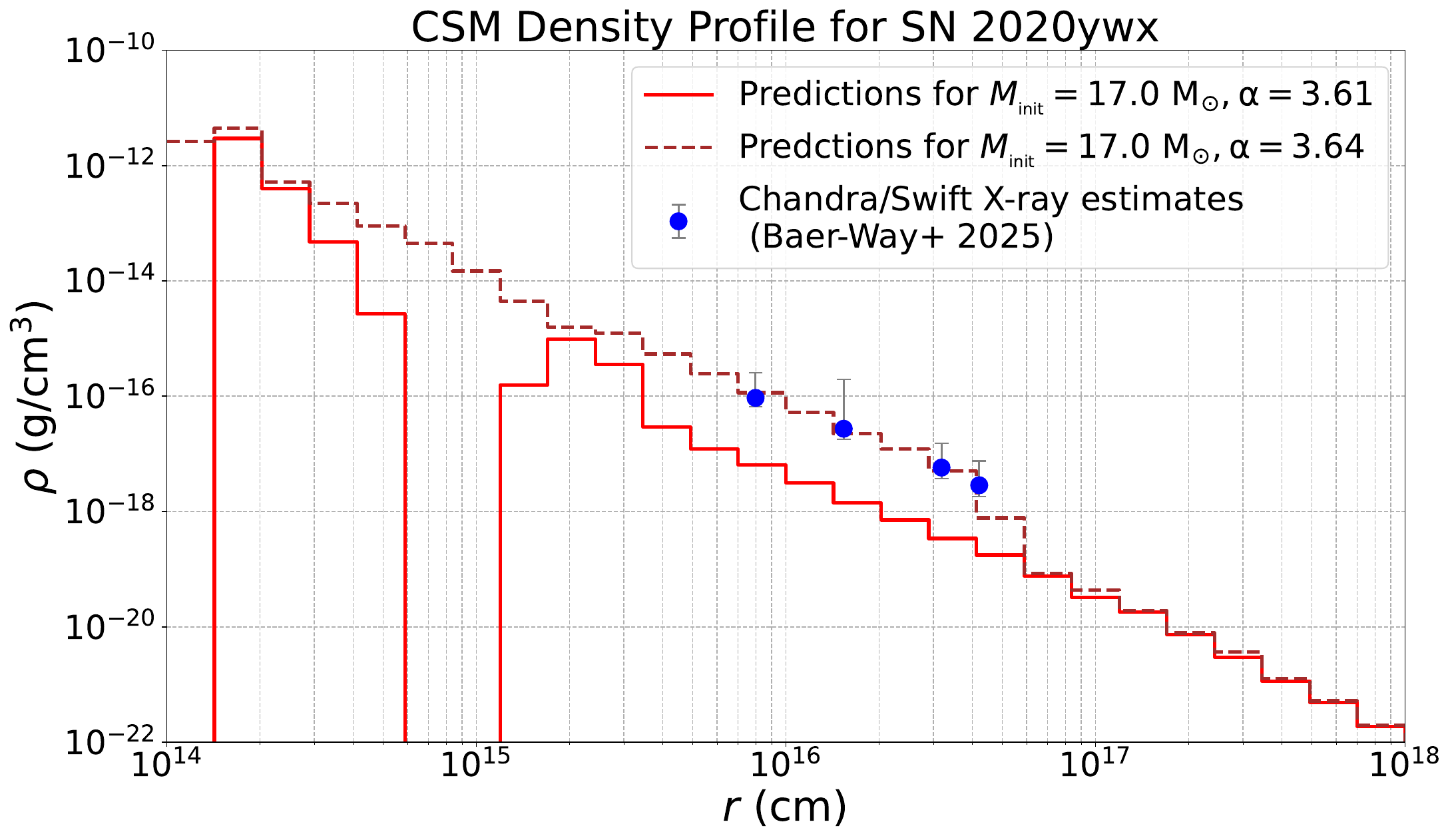}
            \label{fig:9b}
        \end{subfigure}
        \vfill
        \begin{subfigure}[t]{0.99\textwidth}
            \includegraphics[width=\textwidth]{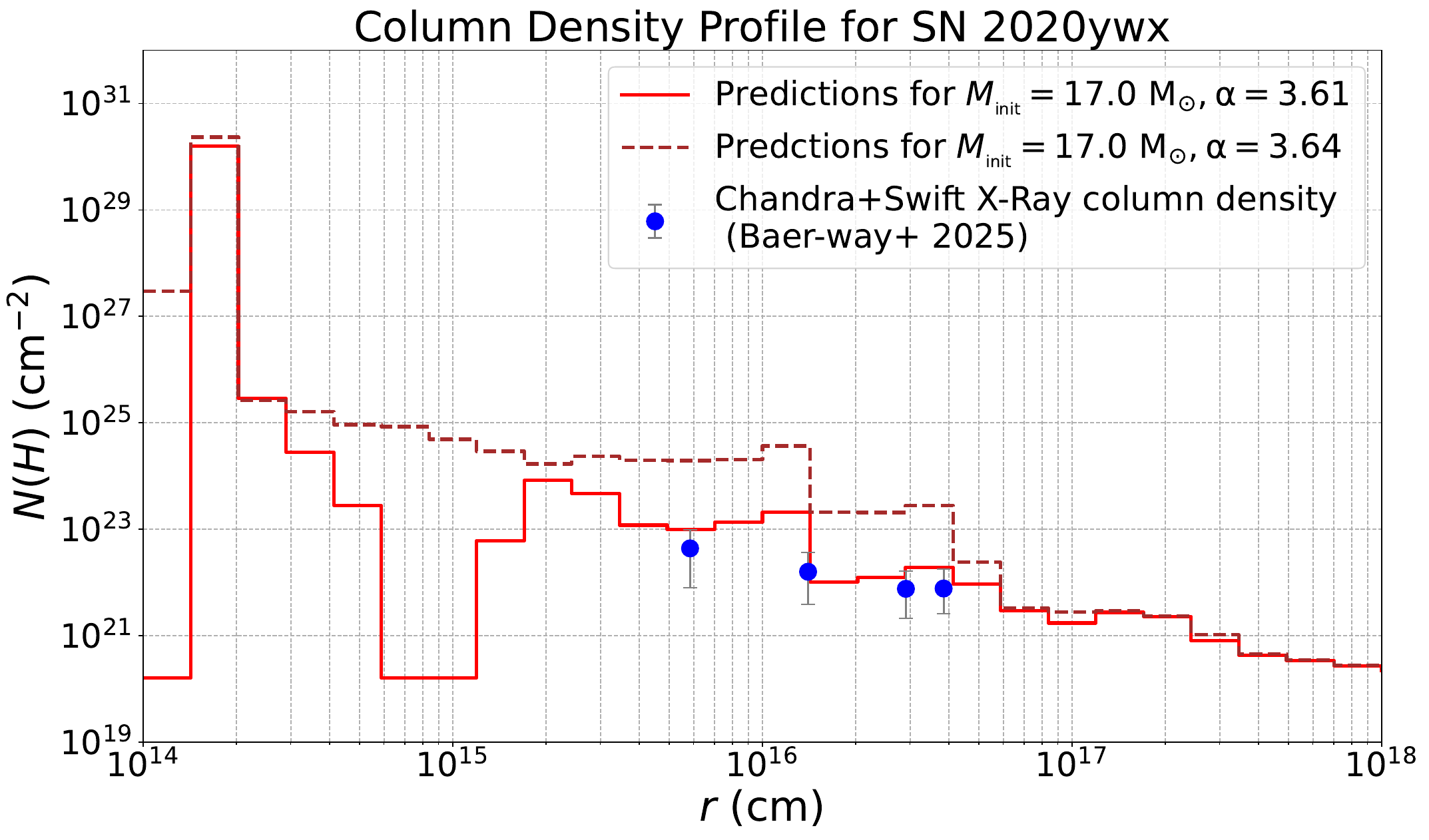}
            \label{fig:9c}
        \end{subfigure}
    
        \caption{
            Evolution of mass-loss rates from our $M_{\rm{init}}=17$~\Ms,~$\alpha=3.61$ and $3.64$ MESA model sequences and the resulting CSM densities from both sequences compared with observationally inferred values along with corresponding column densities of neutral hydrogen compared against measured values for SN-2020ywx:
            \textit{top panel} shows the mass loss histories of the two sequences of our MESA models with $M_{\rm{init}}=17$~\Ms for $~\alpha=3.61$ (in solid red) and $\alpha=3.64$ (in dashed brown), showing enhanced mass-loss rates ($\dot M\gtrsim10^{-2}~$\Ms~$\rm{yr}^{-1}$) from $\sim100$~yrs prior to CC;
            \textit{middle panel} showing corresponding CSM density profiles for the above two MESA model sequences shown in the top panel (the $\alpha=3.61$ sequence is the solid red curve and the $\alpha=3.64$ sequence is the dashed brown curve)  compared with estimates from measured X-ray luminosities for SN 2020ywx \citep{Baer-Way_2025}; \textit{bottom  panel} shows the column densities of neutral hydrogen calculated for the two CSM density profiles shown in the middle panel, compared to the measured X-ray column densities (blue points) from observations \citep{Baer-Way_2025}.
        }
        \label{fig:9}
    \end{minipage}
\end{figure}

SN 2020ywx is the second brightest Type IIn SN observed in X-rays after SN 2010jl, with mass loss rates inferred to be $\gtrsim10^{-2}$~\Ms~$\rm{yr}^{-1}$ over the last $\sim100$~yrs before CC, from multi-epoch Swift and Chandra X-ray observations, while optical and radio data give estimates between $\sim10^{-3}-10^{-2}$~\Ms~$\rm{yr}^{-1}$ \citep{Baer-Way_2025}. To compare with these observed estimates, we present two of our MESA model sequences for $M_{\rm{init}}=17$~\Ms with $\alpha=3.61$ and $3.64$ in the top panel of Figure~\ref{fig:9}. Our $M_{\rm{init}}=17$~\Ms,~$\alpha=3.61$ model sequence exhibits late onset of dynamical ejections with mass loss rates $~10^{-2}$~\Ms~$\rm{yr}^{-1}$ following a superwind phase with lower mass loss rates of $~10^{-4}-10^{-3}$~\Ms~$\rm{yr}^{-1}$, which resembles the optical and radio behaviour of SN 2020ywx. Whereas, our $M_{\rm{init}}=17$~\Ms,~$\alpha=3.64$ model sequence shows a longer phase with dynamical ejections leading to mass loss rates $\gtrsim10^{-2}$~\Ms~$\rm{yr}^{-1}$ over the last $\sim100$~yrs before CC, as is inferred from X-ray luminosities at multiple epochs for SN 2020ywx \citep{Baer-Way_2025}.

The CSM density constructed using the above two model sequences is presented in the middle panel of Figure~\ref{fig:9} along with observational estimates obtained using multi-epoch X-ray mass loss rates from \cite{Baer-Way_2025}, assuming a constant shock velocity of $4000$~km$\rm{s}^{-1}$ and a terminal CSM wind speed of $v_\infty=120\pm22~\rm{km~s}^{-1}$ measured from optical and near-infrared spectra (refer Section~3.1 in \citealt{Baer-Way_2025}). The CSM densities inferred from the observed X-ray mass loss rate estimates are in close agreement with our $M_{\rm{init}}=17$~\Ms,~$\alpha=3.64$ model sequences, while they  are underpredicted by more than an order of magnitude by the $M_{\rm{init}}=17$~\Ms,~$\alpha=3.61$ sequences. This is primarily due to the time of onset and duration of the episodes of dynamical ejections in the two models; also the X-ray mass-loss rates of \cite{Baer-Way_2025} differ by upto an order of magnitude from the optical/radio values beyond the first epoch.

The bottom panel of Figure~\ref{fig:9} presents the column densities of neutral hydrogen along the line of sight for 

\noindent both the models compared to the measured X-ray column densities at the four epochs of Chandra X-ray measurements at $231,445,921~\text{and}~1219$~days of \cite{Baer-Way_2025}. The location of the forward shock within the CSM is found by assuming a constant shock speed of $\sim4000$~km$\rm{s}^{-1}$. It is readily seen that the column densities predicted using our $M_{\rm{init}}=17$~\Ms,~$\alpha=3.61$ model fits the observed values better than the profile obtained using the $M_{\rm{init}}=17$~\Ms,~$\alpha=3.64$ model. The reason for this discrepancy could again arise due to asymmetry effects or line-of-sight effects influencing the column density measurements of \cite{Baer-Way_2025}, as discussed for SN 2023ixf \citep{Chandra_2024,2025ApJ...985...51A}.

\subsection{SN 2017hcc}\label{subsection: 2017hcc}
\begin{figure}[H]
    \begin{minipage}[t]{0.49\textwidth}
        \raggedleft  
        \begin{subfigure}[t]{0.99\textwidth}
            \includegraphics[width=\textwidth]{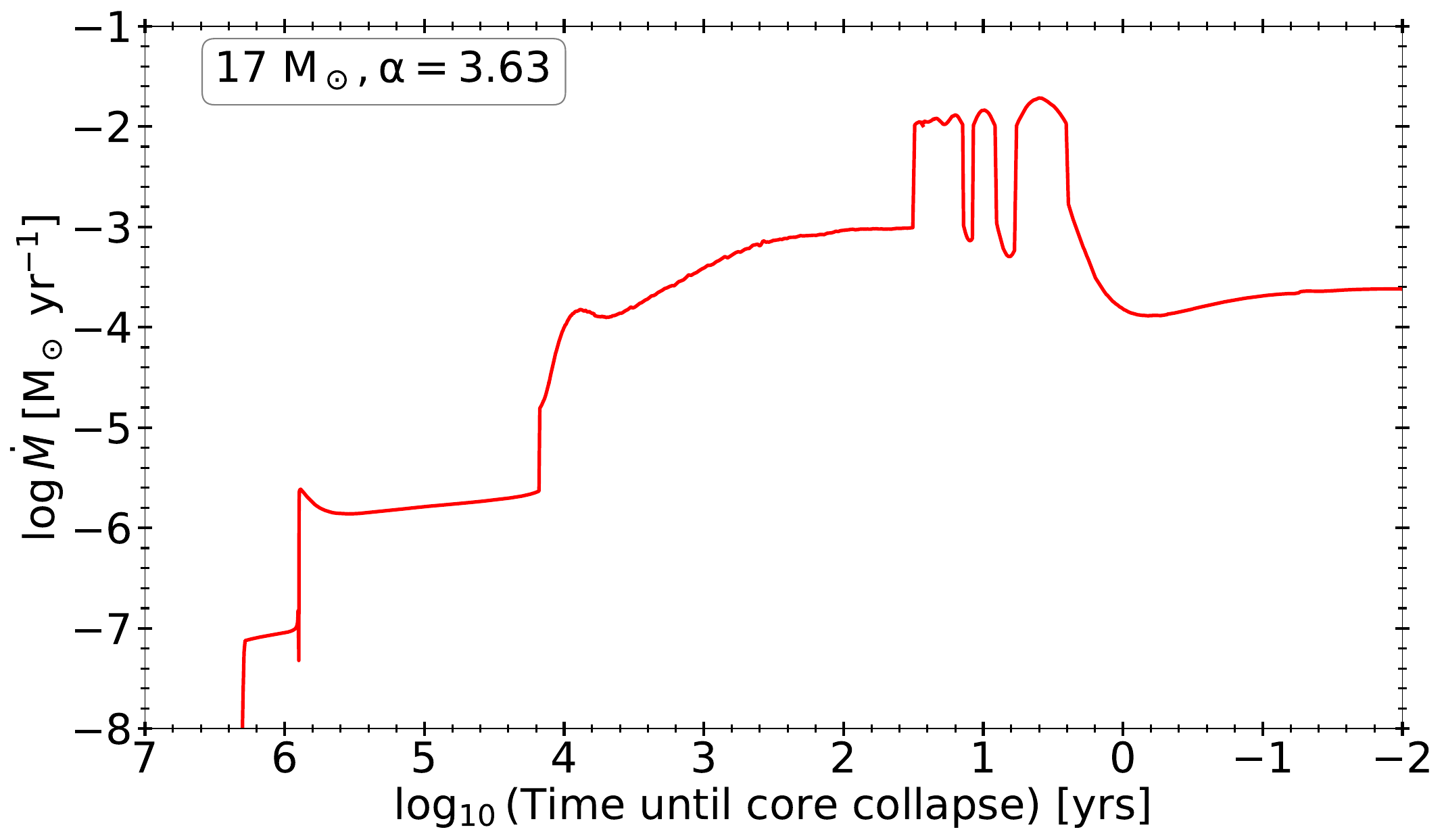}
            \label{fig:8a}
        \end{subfigure}
        \vfill
        \begin{subfigure}[t]{0.99\textwidth}
            \includegraphics[width=\textwidth]{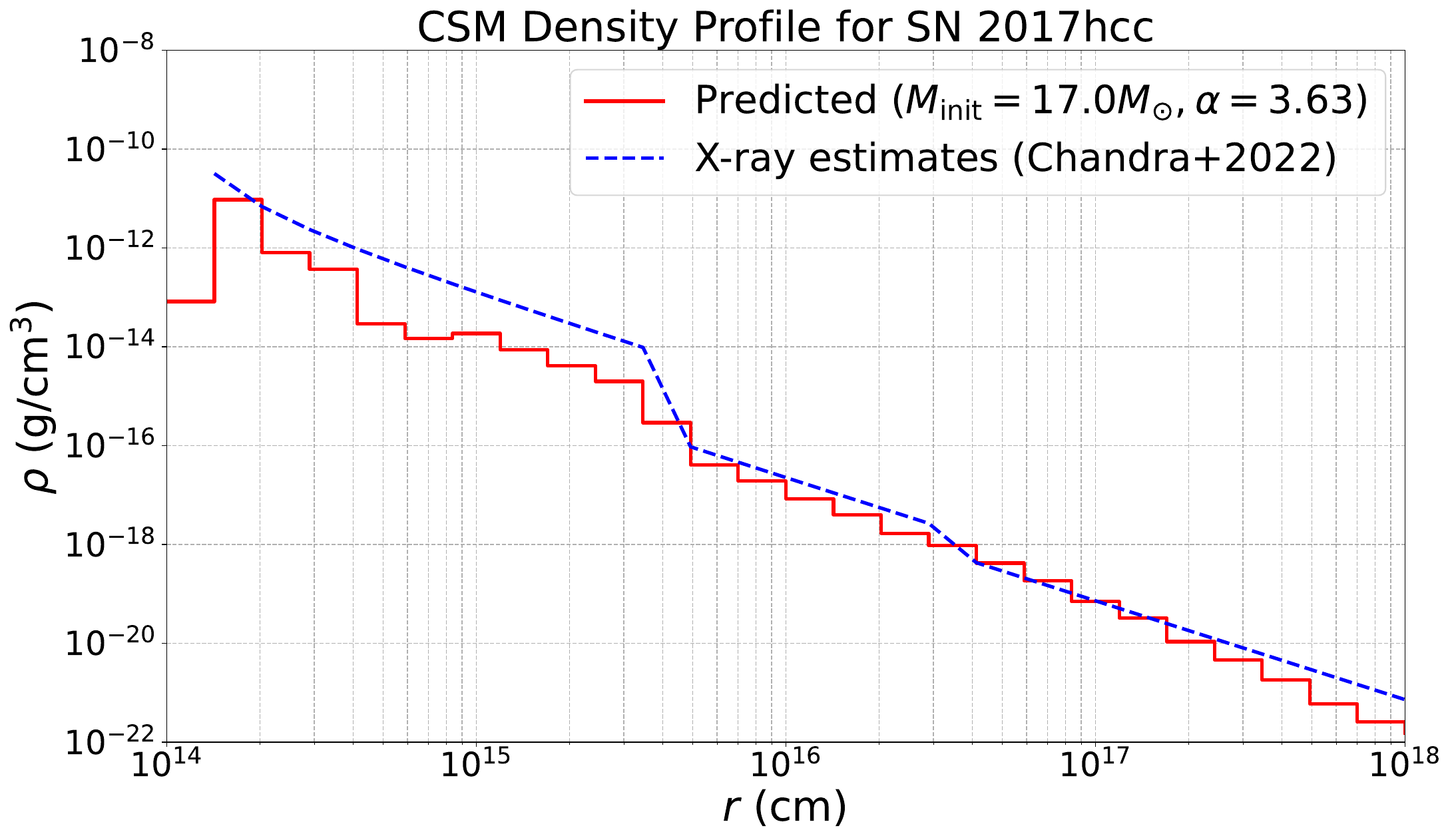}
            \label{fig:8b}
        \end{subfigure}
        \vfill
        \begin{subfigure}[t]{0.99\textwidth}
            \includegraphics[width=\textwidth]{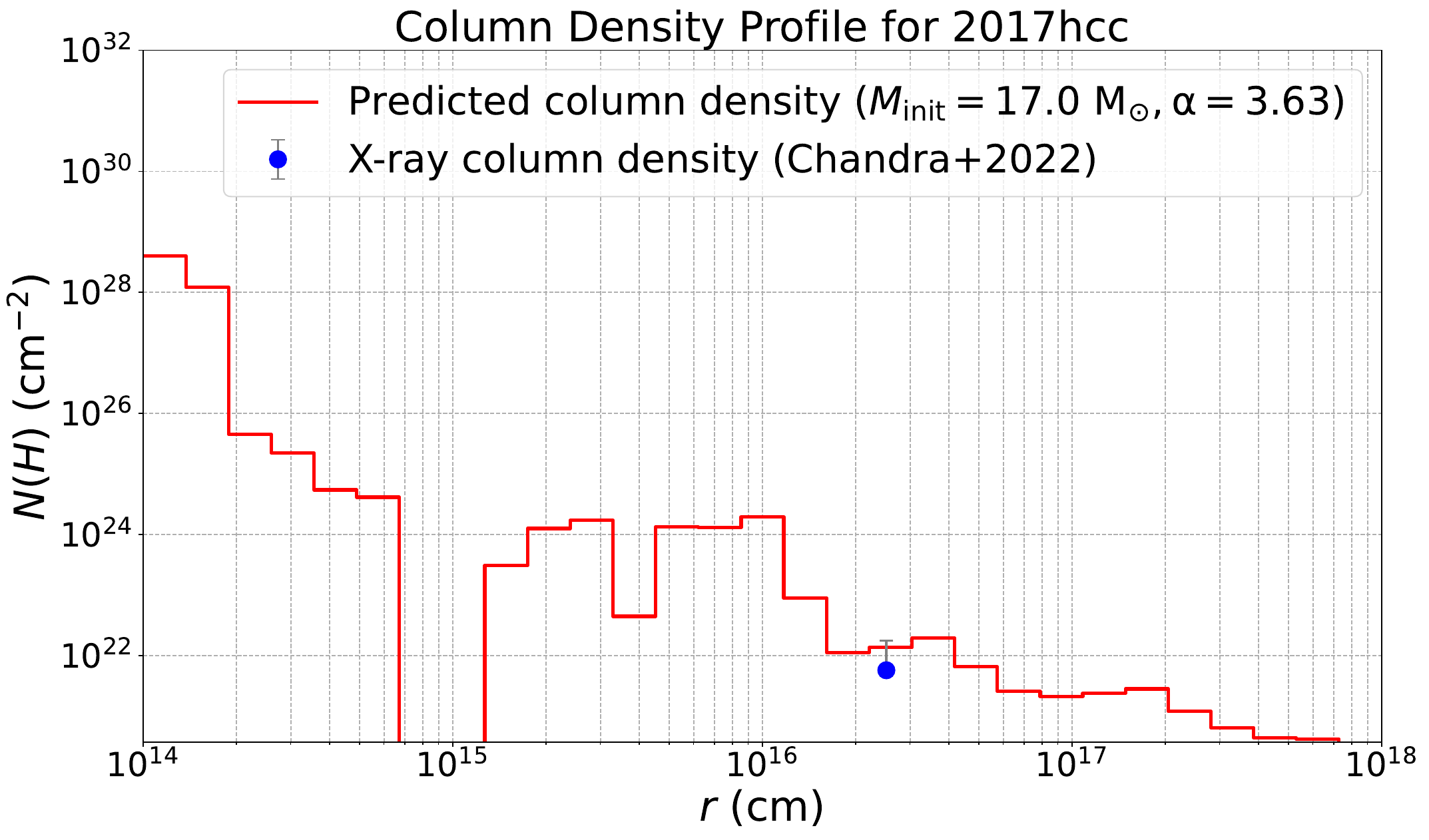}
            \label{fig:8c}
        \end{subfigure}
    
        \caption{
            Evolution of mass-loss rates from our $M_{\rm{init}}=17$~\Ms,~$\alpha=3.63$ MESA model sequence along with the resulting CSM density profile compared with observationally inferred values for SN 2017hcc and the corresponding column densities of neutral hydrogen compared with measured X-ray column density for SN 2017hcc \citep{2022MNRAS.517.4151C}: \textit{top panel} shows the mass loss history of the sequence of MESA models with $M_{\rm{init}}=17$~\Ms,$~\alpha=3.63$, showing mass-loss rate ($\dot M\gtrsim10^{-2}$~\Ms~$\rm{yr}^{-1}$) starting $\sim30$~yrs before CC, due to episodes of dynamical ejections following a superwind phase with $\dot M\gtrsim10^4$~\Ms~$\rm{yr}^{-1}$; \textit{middle panel} shows the CSM density profile for the above MESA model sequence (in solid red), compared with the observationally inferred density profile (in dashed blue) using the mass-loss rate estimates for SN 2017hcc from X-ray observations of \cite{2022MNRAS.517.4151C} as described in \ref{subsection: 2017hcc}; \textit{bottom panel} shows the column densities of neutral hydrogen calculated using the CSM density profiles in the middle panel, compared to the measured X-ray column density (blue point) from Chandra observations \citep{2022MNRAS.517.4151C}.
        }
        \label{fig:8}
    \end{minipage}
\end{figure}

SN 2017hcc belongs to the category of superluminous Type IIn SNe, whose progenitor is thought to have experienced extreme eruptive mass loss in the decade before explosion, with mass loss rate of $\sim1.4$~\Ms~$\rm{yr}^{-1}$ required to power the main peak in its light-curve within first $\sim100$~days dominated by CSM interaction post-explosion \citep{Smith&Andrews2020}. This episodic mass loss is supposed to have occurred for about $6-12$~yrs pre-explosion in order to have created a dense CSM around the progenitor that powered the SN luminosity through strong interaction with this dense CSM. In the top panel of Figure~\ref{fig:8}, we present one of our MESA model sequences with $M_{\rm{init}}=17$~\Ms,~$\alpha=3.63$ which also exhibits episodic mass loss due to dynamical ejections over the last few decades before CC, with mass loss rates reaching above $10^{-2}$~\Ms~$\rm{yr}^{-1}$.

To construct the CSM around the progenitor of SN 2017hcc, we use X-ray observations of \cite{2022MNRAS.517.4151C} who found a lower average mass loss rate of $\sim0.1$~\Ms~$\rm{yr}^{-1}$ over the last decade pre-CC, corresponding to a post shock breakout time of $\sim100$~days, followed by a decline in the mass loss rates to $\sim2\times10^{-3}$~\Ms~$\rm{yr}^{-1}$ at few $100$~days, assuming a bolometric luminosity $\sim10^{42}$~erg$\rm{s}^{-1}$, a terminal wind speed of $v_\infty=45$~$\rm{km~s}^{-1}$ \citep{Smith&Andrews2020} and the velocity of the forward shock to be $4000$~$\rm{km~s}^{-1}$. Furthermore, \cite{2022MNRAS.517.4151C} also present radio observations from which they estimate a mass loss rate at $\sim1000$~days to be $6.5\times10^{-4}$~\Ms~$\rm{yr}^{-1}$. 

The middle panel of Figure~\ref{fig:8} shows a comparison of our model predictions with the CSM densities obtained using the observed mass loss rates from \cite{2022MNRAS.517.4151C} as described above. It can be seen that our model predictions closely trace the observed density profile, even with lower average mass loss rates during both the superwind and dynamical mass ejection phases. While LBV-type eruptive mass loss over the last decade could also  produce a similar density profile \citep{10.1093/mnras/stz1914}, we demonstrate that dynamical ejections following a superwind phase can create a dense enough CSM as is needed to power the peak luminosity of SN 2017hcc.

In the bottom panel of Figure~\ref{fig:8}, we present the neutral hydrogen column density calculated using the predicted CSM density profile shown in the middle panel of  Figure~\ref{fig:8}, along with column density measured from the single epoch of Chandra observations of SN 2017hcc, as presented in \cite{2022MNRAS.517.4151C}. Our model prediction agrees with the observed column density within the uncertainties of its measured value.

\subsection{SN 2005ip}
SN 2005ip is the one of the best studied Type IIn SN with a long-lasting phase of strong interaction with its CSM hinting towards high mass loss rates ($\gtrsim10^{-2}$~\Ms~$\rm{yr}^{-1}$) over the last $\sim10^{2}$~yrs before CC \citep{Fox2020}. Eruptive mass loss from LBVs have been invoked in the literature to explain the origin of the dense shell around SN 2005ip \citep{Smith_2009,Katsuda_2014}. Here, we illustrate that dynamical ejections following a strong superwind phase from pulsating RSGs could also lead to enduring phases of such heavy mass loss, as is needed to reconcile with measured column densities from Swift and Chandra X-ray observations of SN 2005ip \citep{Katsuda_2014, Fox2020}.

\begin{figure}[H]
    \begin{minipage}[t]{0.49\textwidth}
        \raggedleft  
        \begin{subfigure}[t]{0.99\textwidth}
            \includegraphics[width=\textwidth]{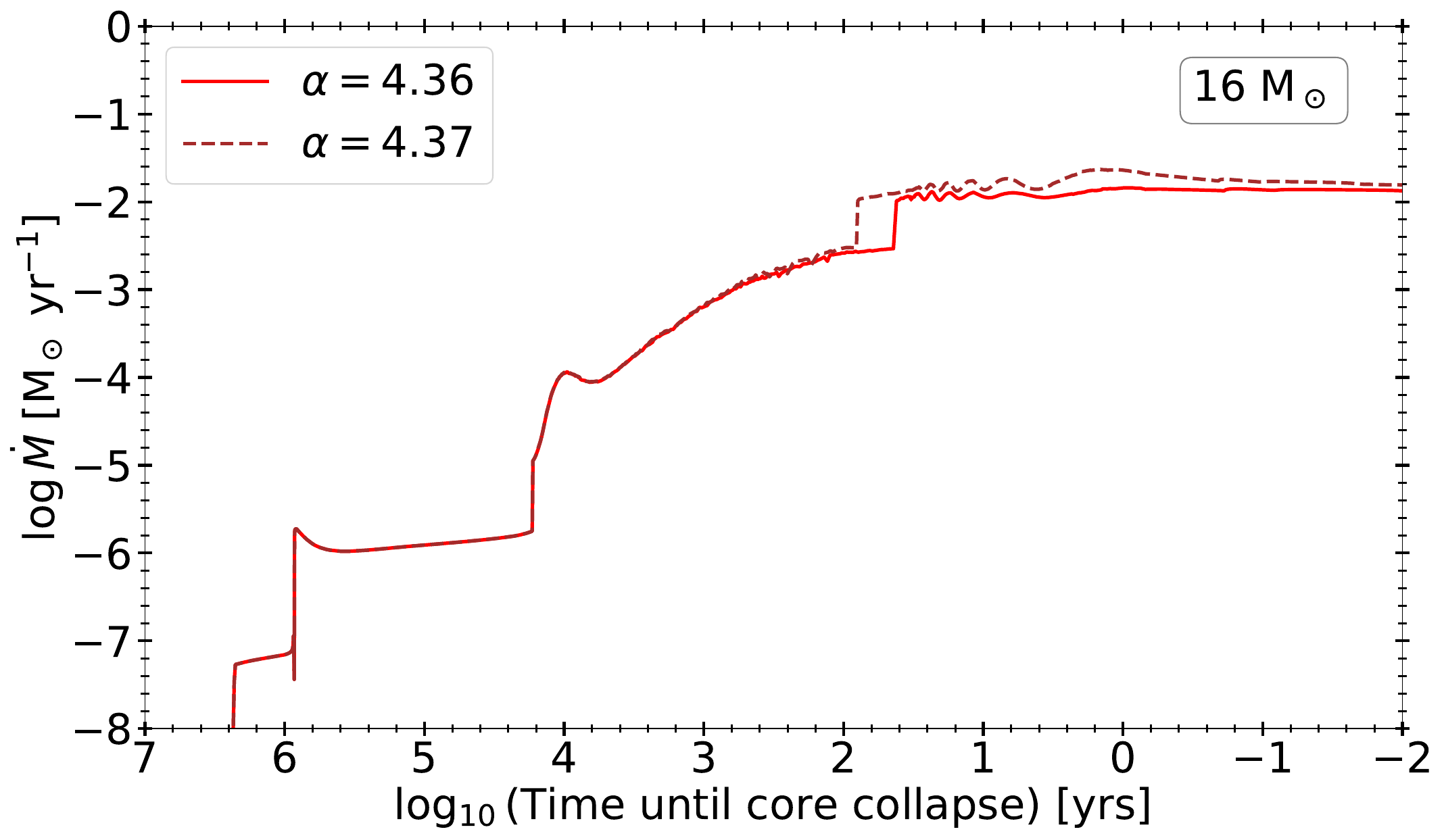}
            \label{fig:7a}
        \end{subfigure}
        \vfill
        \begin{subfigure}[t]{0.99\textwidth}
            \includegraphics[width=\textwidth]{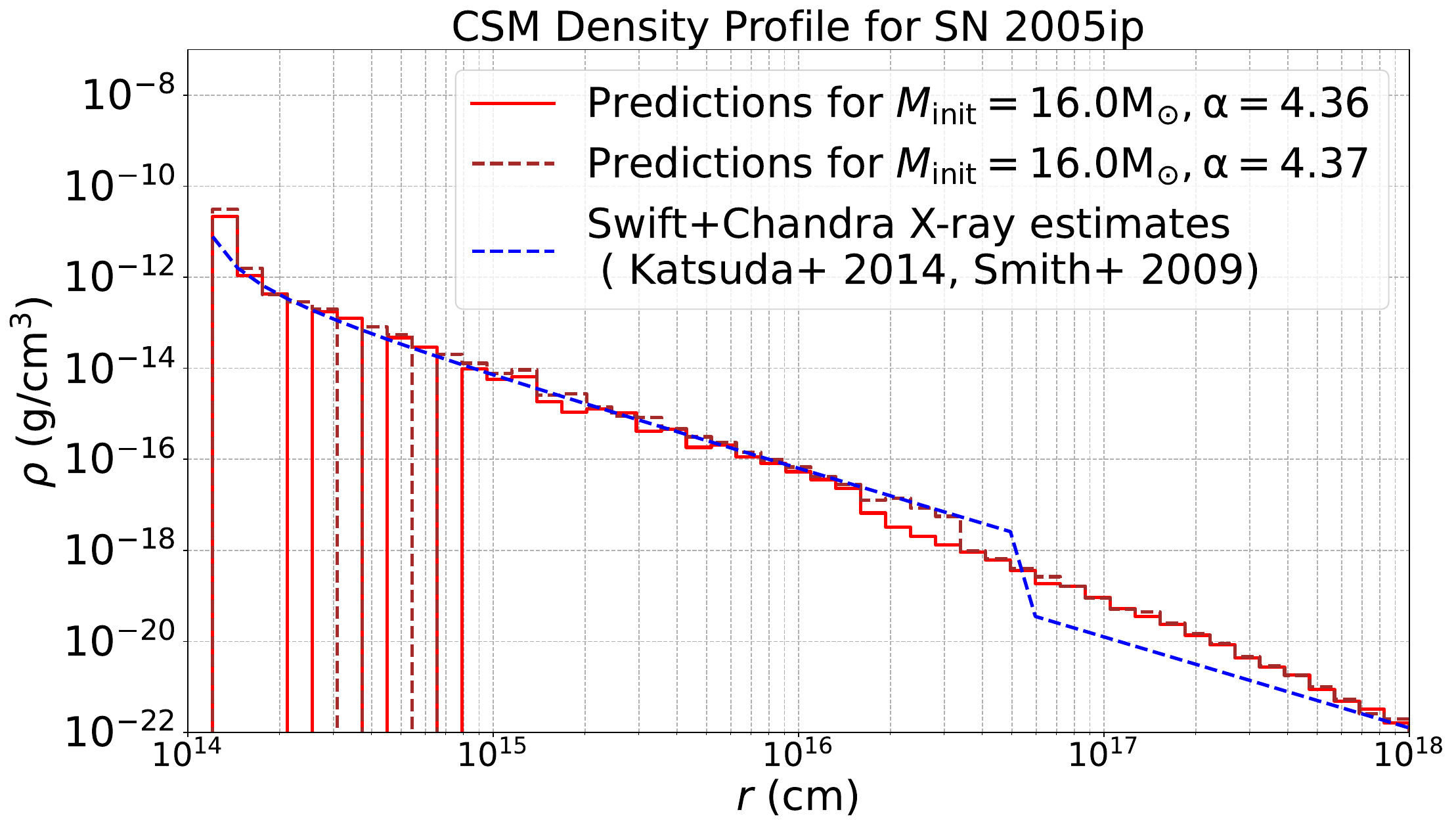}
            \label{fig:7b}
        \end{subfigure}
        \vfill
        \begin{subfigure}[t]{0.99\textwidth}
            \includegraphics[width=\textwidth]{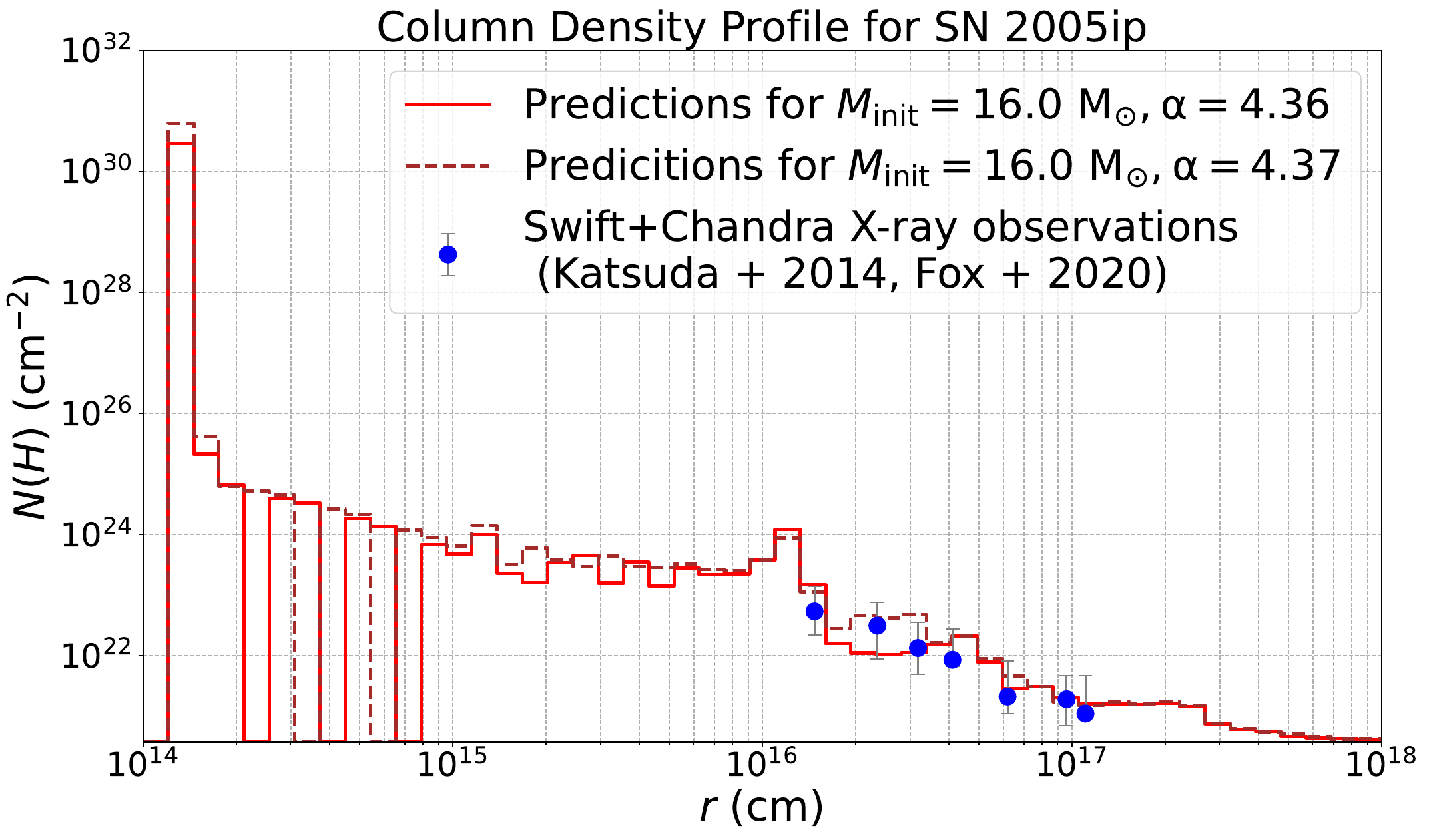}
            \label{fig:7c}
        \end{subfigure}
    
        \caption{
            Evolution of mass-loss rates from our $M_{\rm{init}}=16$~\Ms,~$\alpha=4.36$ and $4.37$ MESA model sequences with resulting CSM densities from both sequences compared with observationally inferred values along with corresponding column densities of neutral hydrogen compared against measured values for SN 2005ip:
            \textit{top panel} shows the mass loss histories of the two sequences of MESA models with $M_{\rm{init}}=16$~\Ms for $~\alpha=4.36$ (in solid red) and $\alpha=4.37$ (in dashed brown), showing enhanced mass-loss rates ($\dot M\gtrsim10^{-2}~$\Ms~$\rm{yr}^{-1}$) starting $\sim44$ and $\sim81$~yrs respectively before CC; \textit{middle panel} shows the CSM density profiles for the above two MESA model sequences in the top panel (the $\alpha=4.36$ sequence is the solid red curve and the $\alpha=4.37$ sequence is the dashed brown curve), compared with observationally inferred profile (in dashed blue) for SN 2005ip; \textit{bottom panel} shows the column densities of neutral hydrogen corresponding to the CSM density profiles shown in middle panel, compared to the measured  X-ray column densities (in solid blue) from observations \citep{Katsuda_2014, Fox2020}.
        }
        \label{fig:7}
    \end{minipage}
\end{figure}

The top panel of Figure~\ref{fig:7} shows two sequences of MESA models both with $M_\text{init}=16$~\Ms, but with $\alpha=4.36~\rm{and}~4.37$ exhibiting such a sustained period of enhanced mass loss from dynamical ejections following a strong superwind phase with lower mass loss rates between $10^{-4}-10^{-3}$~\Ms~$\rm{yr}^{-1}$. The dynamical ejections occur with an average mass loss rate $\gtrsim10^{-2}$~\Ms~$\rm{yr}^{-1}$ over several decades before CC, which leads to $\gtrsim1$\Ms of CSM material located close to the  progenitor, as seen in the middle panel of Figure~\ref{fig:7}. Our model predictions closely trace the observationally inferred CSM density, computed using a mass loss rate of $1.5\times10^{-2}$~\Ms~$\rm{yr}^{-1}$ \citep{Katsuda_2014} and a terminal wind speed of $v_\infty=120$~$\rm{km~s}^{-1}$ \citep{Smith_2009}, out to  distances of $\sim5\times10^{16}$~cm from the progenitor surface which is the location of the forward shock at $\sim2000$~days post-explosion, assuming the shock velocity $V_{\rm{SH}}\sim t^{-0.1}$ and has a value of $3000~\rm{kms}^{-1}$ at $t=1000$~days \citep{Fox2020}. At later times, corresponding to larger distances,  the mass loss rates are estimated to be much lower at $\sim(2-4)\times10^{-4}$~\Ms~$\rm{yr}^{-1}$ \citep{Smith_2009}, which is also reflected in our model densities caused due

\noindent to matter lost in the superwind phase preceding dynamical ejections. The model with $\alpha = 4.37$, which has an earlier onset of dynamical ejections at $~80$~yrs from CC, shows this drop in CSM density at a location closer to the observed extent of the dense shell around the progenitor \citep{Fox2020}.

The bottom panel of Figure~\ref{fig:7} \citep{Smith_2009}, shows the column density profile of neutral hydrogen along the line of sight obtained using the CSM densities according to \ref{subsection :  col density}. The calculated column densities from the model with $\alpha=4.36$ are in close agreement with available estimates from late-time Swift/Chandra X-ray observations of \cite{Katsuda_2014} and \cite{Fox2020}, which are converted into radial coordinates from post-explosion times using the time evolution of the shock wavefront, as mentioned above.

\subsection{SN 1998S}
SN 1998S is considered to be the prototypical Type-IIn SN, with a mass loss rate estimate of $6\times10^{-3}$~\Ms~$\rm{yr}^{-1}$ during the last $\sim15$~yrs before CC and a much lower rate at earlier times \citep{2015ApJ...806..213S}, though \cite{2012MNRAS.424.2659M}  showed substantial matter present as far out as $r\sim2\times10^{17}$~cm from late-time spectroscopic observations $14$~yrs post-explosion. Previously, the terminal velocity of the wind and the mass loss rate of the progenitor had been estimated to be $40-50$~km/s and $(1-2)\mathbf{\times10^{-4}}$~\Ms~$\rm{yr}^{-1}$ respectively \citep{2001MNRAS.325..907F,Pooley_2002}. Based on these studies, we use the following observed mass loss rate profile:   
\begin{equation} \label{equation : 8}
\dot{M}_\text{obs} =
    \begin{cases}
    6\times10^{-3}~\rm{M}_\odot~\rm{yr}^{-1}, & r<10^{15}~\text{cm} \\
    2\times10^{-4}~\rm{M}_\odot~\rm{yr}^{-1}, & r>10^{15}~\text{cm},
    \end{cases}
\end{equation}
along with a terminal wind speed $v_{\infty}=40$~$\rm{km~s}^{-1}$,~$v_0=0.1$~km/s and $R_*=1.1\times10^{14}~cm$ to calculate the observationally inferred CSM density as per Equation \ref{equation : 5}.
\begin{figure}[ht]
    \centering
    \begin{subfigure}[t]{0.45\textwidth}
        \includegraphics[width=\textwidth]{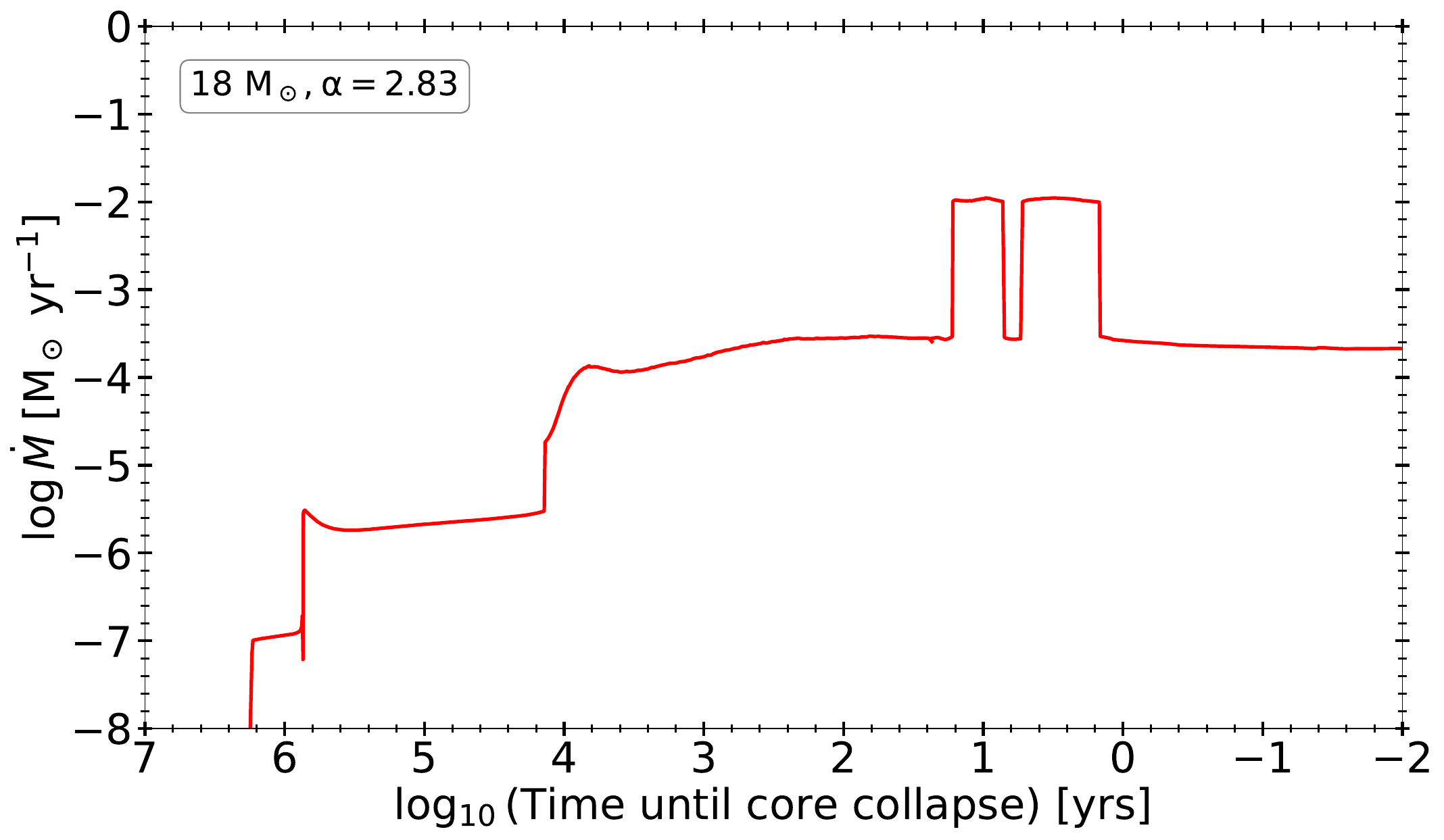}
        \label{fig:6a}
    \end{subfigure}
    \hfill
    \begin{subfigure}[t]{0.45\textwidth}
        \includegraphics[width=\textwidth]{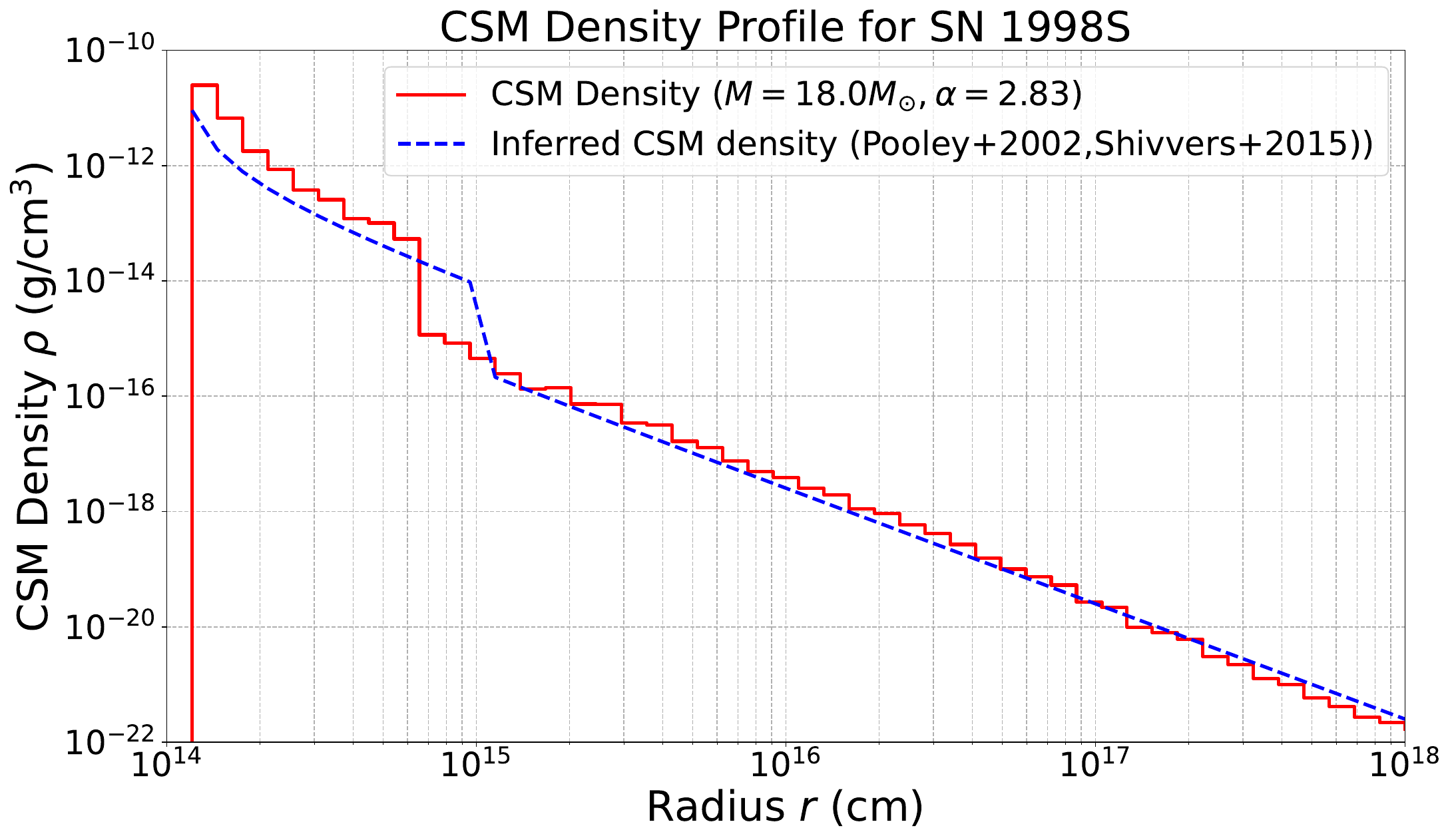}
        \label{fig:6b}
    \end{subfigure}
    
    \caption{
        Evolution of mass-loss rates for  $M_{\rm{init}}=18$~\Ms,$~\alpha=2.83$~ MESA model sequences along with corresponding CSM density compared with observationally inferred estimates for SN 1998S: \textit{top panel} shows the mass loss history of the sequence of MESA models with $M_{\rm{init}}=18$~\Ms,$~\alpha=2.83$~prior to CC, showing enhanced mass-loss rates ($\dot M\gtrsim10^{-4}$~\Ms~$\rm{yr}^{-1}$) starting $\sim10^{4}$~yrs, due to PDSWs, followed by the onset of dynamical ejections at $\sim16$~yrs from CC (\ref{subsection : deso}); \textit{bottom panel} shows the CSM density profile (in solid red) computed for the adjacent sequence of MESA models according to Equation~\ref{equation : 5}, compared with the observationally inferred estimate (in dashed blue) using the mass-loss profile given by Equation~\ref{equation : 8} for SN 1998S.
    }
    \label{fig:6}
\end{figure}

In top panel of Figure~\ref{fig:6}, we present our $18$~\Ms,~$\alpha=2.83$ MESA model sequences which shows enhanced mass loss rates ($\dot M>10^{-4}$~\Ms~$\rm{yr}^{-1}$) due to superwinds from $\sim10^{3.5}$~yrs up to $\sim16.22$~yrs before CC, after which there is a sudden increase in mass loss rate ($\sim10^{-2}$~\Ms~$\rm{yr}^{-1}$) from episodes of dynamical ejections (as described in Section~\ref{subsection : deso}), until $\sim1.48$~yr prior to CC. These properties are similar to the observed estimates for the mass loss of the progenitor of SN 1998S as discussed above, with mass loss rates as given by Equation~\ref{equation : 8} within a factor of $\sim2$ of those in these MESA model sequences.

The bottom panel of Figure~\ref{fig:6} compares the CSM density predicted from the MESA models using Equation~\ref{equation : 5} with the values inferred from the observed mass loss rate profile given by Equation~\eqref{equation : 8}. The "inner" CSM ($r<10^{15}$~cm) predictions closely resemble the observationally estimated density profile for $r\lesssim7\times10^{14}~\mathrm{cm}$. Other evolutionary scenarios invoked to explain the observed light curve of SN 1998S, which requires a dense CSM nearby, include merger of a massive RSG with a compact object \citep{2020ApJ...892...13S}. However, the presence of two distinct shells as described above poses tension with this merger-driven explosion scenario, though episodic mass ejections may still occur during the common-envelope (CE) phase in such systems \citep{2017MNRAS.470.1788C}. 

\section{Discussions}\label{section: discussions}

\subsection{Observational implications for the RSG problem}
Our MESA evolutionary sequences with the PDSW and subsequent dynamical ejections for initial masses $\lesssim18$~\Ms,   presented in Section~\ref{section: pulsating RSGs}, undergo late-stage heavy mass loss, creating the dense CSM around the progenitor star as is required to explain observations of the Type II SNe shown in Section~\ref{section : comparison}. \cite{2010ApJ...717L..62Y} suggested that, considering their PDSW mass loss prescription, their exists a critical mass, $M_{\rm{crit}} \sim 19-20$~\Ms~above which RSGs lose a large fraction of their H-rich envelopes prior to CC, without producing Type IIP SNe, reconciling with the so-called missing RSG problem inferred from observations of RSGs \citep{2009MNRAS.395.1409S,2015PASA...32...16S,davies_2020,kochanek_2020,beasor_2025,2025ApJ...986...39F}. We performed some calculations with higher initial mass in the above range, most of which led to runaway mass loss due to shock-driven mass ejections shortly following the onset of PDSW, in the very early stages of core-carbon burning. We could not evolve these runs further due to numerical convergence issues, but such runaway events also favor the scenario in which most of the H-rich envelope lost much before CC, in agreement with the observed lack of higher mass RSGs associated with Type IIP SNe. 

\subsection{Pulsation-driven mass loss in binary systems}
Large-amplitude pulsations can also drive enhanced mass loss in binary systems, both through PDSWs of the kind considered in this study, as well as mass transfer via pulsational Roche lobe overflow \citep{2007ASPC..372..397M}, by changing the Roche lobe radius and its impact on mass transfer in the presence of pulsation-driven winds \citep{2009A&A...507..891D}. \cite{2017MNRAS.470.1788C} also investigated the possibility of dynamical mass ejections during the inspiral of a compact object within the common envelope (CE) of a merging binary system, demonstrating the viability of envelope ejection following episodes of strong compression. Such scenarios may also have profound impacts on the binary evolution outcomes of more massive stars -- the majority of which are thought to have experienced at least one episode of interaction with a companion star during the course of their evolution \citep{2012Sci...337..444S}. Hence, a systematic study of the effects of both pulsation-driven mass loss and shock-driven dynamical mass ejections is warranted in the context of massive binaries to quantify the impact on pre-SN evolution as well as CSM formation. 

Incorporating these mass loss mechanisms might also help to alleviate the existing tension between pre-explosion mass loss rates inferred from observations of Type Ib/c SNe \citep{2016ApJ...821...57D} and those measured from stripped progenitor stars in binaries \citep{2023ApJ...959..125G}. Furthermore, it has been proposed late mass-transfer episodes in binaries may be stable if they are pulsed, possibly as a consequence of pulsation instabilities in the envelopes of the giant donors in such systems \citep{10.1007/978-94-009-2413-0_15} (for detailed estimates of all probable massive binary evolutionary channels leading to various types of SNe, refer \citealt{1992ApJ...391..246P}).    

\subsection{Effects of rotation and metallicity}
The effect of rotation on the pulsational instability of RSG envelopes was examined by \cite{1997A&A...327..224H} who found dramatic increases in $L/M$ ratios due to the physical effects of rotation on the evolution of massive stars, even at typical values of surface rotational velocities of $\sim200~kms^{-1}$ which amounts to $\lesssim35 \%$ of critical rotation. This effect is primarily due to an increase in helium core masses in the evolutionary models, which translates to higher luminosities during advanced phases of evolution. This property leads to a significant increase in the growth rates of the fundamental mode of the pulsational instability, which in turn would lead to further enhancement in the strength of the associated PDSW (as per Equation~\ref{equation : 2}). \cite{2010ApJ...717L..62Y} concluded that rotation could further lower the effective upper mass limit ($M_{\rm{crit}}$) of RSGs which are expected to explode as Type IIPs, in closer agreement with latest studies on the observed value of this limit \citep{beasor_2022,beasor_2025}. This would also have consequences for the formation of Wolf-Rayet (WR) stars that are thought to be single star progenitors of Type Ib/c SNe \citep{1996ASPC...98..220L}, as more rapidly rotating RSGs might already show strong pulsations during core helium burning and lose sufficient mass to evolve into WR stars \citep{rsg_wr2012}. It would also imply lower initial masses for WR stars than predicted from standard non-rotating evolutionary models, thereby increasing the number of single star progenitors of Type Ib/c SNe \citep{1993ApJ...411..823W}.

Metallicity also plays a key role in driving mass loss through line-driven winds \citep{Heger_2005,refId0,2012A&A...546A..42M}, with weaker winds predicted at lower metallicities (\citealt{1991A&A...245..548B,2025A&A...699A..71H}; though see \citealt{2011A&A...526A.156M,2025arXiv250305876A}) due to lack of metal lines needed to absorb sufficient number of photons which can impart necessary momentum to drive the wind. Due to this effect, it has been proposed that massive stars only evolve into cool RSGs above a critical metallicity \citep{2023ApJ...944...34O}, below which they can advance through core helium or carbon burning while retaining most of their compact envelopes \citep{2025arXiv250601753O}. \cite{2001ApJ...550..890B} studied the pulsational stability of very massive stars of varying metallicity and found a transition for the mechanism driving these pulsations with increasing metallicity, with the $\kappa$-mechanism dominating for metal-rich counterparts of primordial stars that may end their lives without losing much mass \citep{2010Ap&SS.327..219B}. This is attributed to the much longer growth timescales of nuclear-powered pulsational instabilities in metal-free stars on the main sequence, which quickly stabilize after evolving off the main sequence. However, \cite{ml2015} demonstrated that pulsationally driven winds could be an efficient way for very massive stars to lose a substantial part of the envelopes in the RSG phase before explosion, even at zero metallicity. Furthermore, the prospect of dynamical mass loss following subsequent shock breakout in the outer layers of envelopes of RSGs in such pristine regimes, also deserves attention, particularly in the context of dust formation in the high-redshift universe \citep{2015ComAC...2....3G} as observed by large-scale surveys such as the JWST and the LSST.

\subsection{Implications for pre-explosion dust formation around RSGs}

RSG stars are often associated with a dusty circumstellar medium \citep{walmswell_2012, clayton_2009, cannon_2021, niu_2023, gaston_2025}. Yet the modest mass loss rates ($\sim$ 10$^{-6}$~\Ms) characterizing RSG winds are found unsuitable to synthesize dust grains from the gas-phase atoms \citep{cherchneff2013b}. Dust formation requires moderately low temperatures while the gas densities are still high to allow series of collisions leading to the formation of gas-phase precursors \citep{sarangi2018book}. In the RSG with a dense CSM, the enhanced mass loss can aid the formation of chemical compounds in the wind, depending on the wind velocities. As we have shown, each stellar model evolves in unique routes, which will result in a unique distribution of molecules and dust grains. If the enhanced mass loss is triggered during the He-burning phase or early C-burning phase, the dust formed in that gas will fly far from the star by the time the star goes to SN. This dust may already integrate with the ambient interstellar medium (ISM), and enhance the dust budget in the galaxy. Moreover, given the large distance, they are shielded from the strong ionizing SN flashwave at the point of core-collapse. However, this dust is difficult to detect in stellar spectroscopy in optical or IR, given it will be at a large distance (hence causing no significant extinction to the stellar light curve) and cold temperatures (making it indistinguishable with the ambient ISM). 

On the other hand, in the RSGs with enhanced mass loss at the very end of their lives, the dusty CSM significantly impacts the stellar light curve and its post-explosion IR signatures. In addition, the composition of the CSM may also be altered based on the nature of the stellar evolution and mass loss. Even though the CSM is often identical in composition with the H-rich stellar envelope, a small variation in the abundance of the metals will impact the composition and dust masses. The formation of dust precursors in the gas is often controlled by some bottleneck gas-phase reactions \citep{sar13}. On the contrary, the growth of grains, which decides the grain size distributions, is proportional to time \citep{hirashita_2012}. The population of larger grains are more resilient to destruction processes post SN explosion, or shocks in the ISM \citep{micelotta2018}. In that regard, the timing of the enhanced mass loss plays key role in determining the grain sizes. 

Our follow-up study is on the formation of dust in the realistic winds of the RSGs that lead to the SNe we have modeled in this paper. 


\subsubsection{Numerical and computational challenges in obtaining pre-SN yields}
In order to predict pre-explosion yields to obtain dust masses, our MESA runs need to be evolved through explosive oxygen and silicon burning that occurs on day-hour timescales. As already discussed, accurate energy generation requires the use of large nuclear networks (e.g. \texttt{mesa\_128.net}) with implicit hydrodynamics and $\sim 10^{5}$ timesteps causing severe slowdowns (with a single run requiring more than a week on a $32$-core CPU machine), making a similar systematic exploration of our parameter space in $M_{\rm{init}}$ and $\alpha$, prohibitively expensive. A hybrid computing strategy -- involving zone-by-zone parallelism on a graphics processing unit (GPU) by offloading the nuclear network calculations that form a major bottleneck of MESA with large nuclear networks -- seems a promising approach to significantly reduce the large computational cost of evolving our runs through to CC. With an initial grid of such models, the recently proposed Nuclear Neural Network (NNN) framework of \cite{Grichener_2025} can be employed to further refine the grid that seems essential to study stochastic effects of convection, shell merging etc. on final yields and associated dust masses.

\section{Conclusions}
This work presents a comprehensive, end-to-end investigation of how pulsating RSGs can create a dense, possibly dusty, CSM through enhanced mass loss episodes in the final years before CC, which shapes the multi-wavelength appearance of the ensuing SN explosion. From a dense grid of MESA models of varying $M_{\rm{init}}=12-20$~\Ms~ and PDSW strength (in terms of the superwind parameter $\alpha=2-5$, as in \citealt{2010ApJ...717L..62Y}), as well as subsequent shock-driven dynamical ejections \citep{clayton2018a}, the mass loss histories of the model sequences are coupled to an accelerated $\beta$-law wind profile to construct spherically symmetric CSM density and column-density profiles that are directly compared to multi-epoch observations of SNe 1998S, 2005ip, 2017hcc, 2020ywx, and 2023ixf. We find decent agreement (within a factor of few) between the observationally inferred values and our model results for $M_{\rm{init}} > 15$~\Ms~ and $\alpha > 2$ which give enhanced mass loss rates ($\dot M \sim 10^{-4}-10^{-2}$~\Ms~yr$^{-1}$) in the final decades to centuries before explosion. Such late phase heavy mass loss produces sufficiently dense CSM close to the SN progenitor to explain early-time observations (e.g. using flash-spectroscopy, X-rays and radio wavelengths) as well as late epoch estimates of X-ray column densities (Section~\ref{section : comparison}). Our results quantitatively demonstrate that single star evolutionary pathways with pulsation-driven mass loss and consequent dynamical ejections can account for the observed diversity in the properties of Type II SN environments.

In future, missing physics such as effects of varying rotation and  metallicity as well as binary interaction needs to be incorporated to improve the realism of our models, all of which are expected to have broader implications both for massive star evolution as well as their final fates in terms of the nature of the eventual SN. Also, the pulsation properties of other classes of massive stars such as WR stars \citep{2016A&A...590A..12G,2018A&A...614A..86G} and LBVs \citep{1999LNP...523..337G,2014ApJ...796..121J} deserve investigation as they are supposed to be prime candidates for progenitors of both Type Ib/c and superluminous SNe. In addition, further improvements in our understanding of the interplay of radial pulsation and convective motions in RSG envelope through multidimensional simulations \citep{2022ApJ...929..156G,2025A&A...703A..61B,2025arXiv251014875M} and its role in driving mass loss are critical towards advances in the predictive power of evolutionary models by incorporation self-consistent treatment of stellar winds through physical mechanisms such as pulsations and shocks, as compared to the empirical approach adopted in this study and extant works in the literature \citep{2010ApJ...717L..62Y,ml2015,2025A&A...703A..61B,2025MNRAS.543.3929S}. This will have important implications for SN modeling at large and their impact on cosmic evolution and dust formation ranging from our local environment to the high redshift universe in the era of large-scale surveys like the JWST and the LSST. 

\section{Acknowledgments}
The authors acknowledge the support of the Department of Science and Technology (DST), Government of India. We thank Dr.~Heloise Stevance and Professor Jan Eldridge for constructive suggestions and helpful comments on this work, and Mr.~Raphael Baer-Way for private communication and for sharing data from \cite{Baer-Way_2025}. This research has made use of the High Performance Computing (HPC) resources\footnote{\url{https://www.iiap.res.in/centers/main-campus/computing-and-i-t/}} made available by the Computer Center of the Indian Institute of Astrophysics (IIA), Bangalore.

\appendix

\section{Choice of $\beta$ parameter for accelerated wind}\label{beta-law}
\begin{figure}[h]
    \centering
    \includegraphics[width=0.45\textwidth]{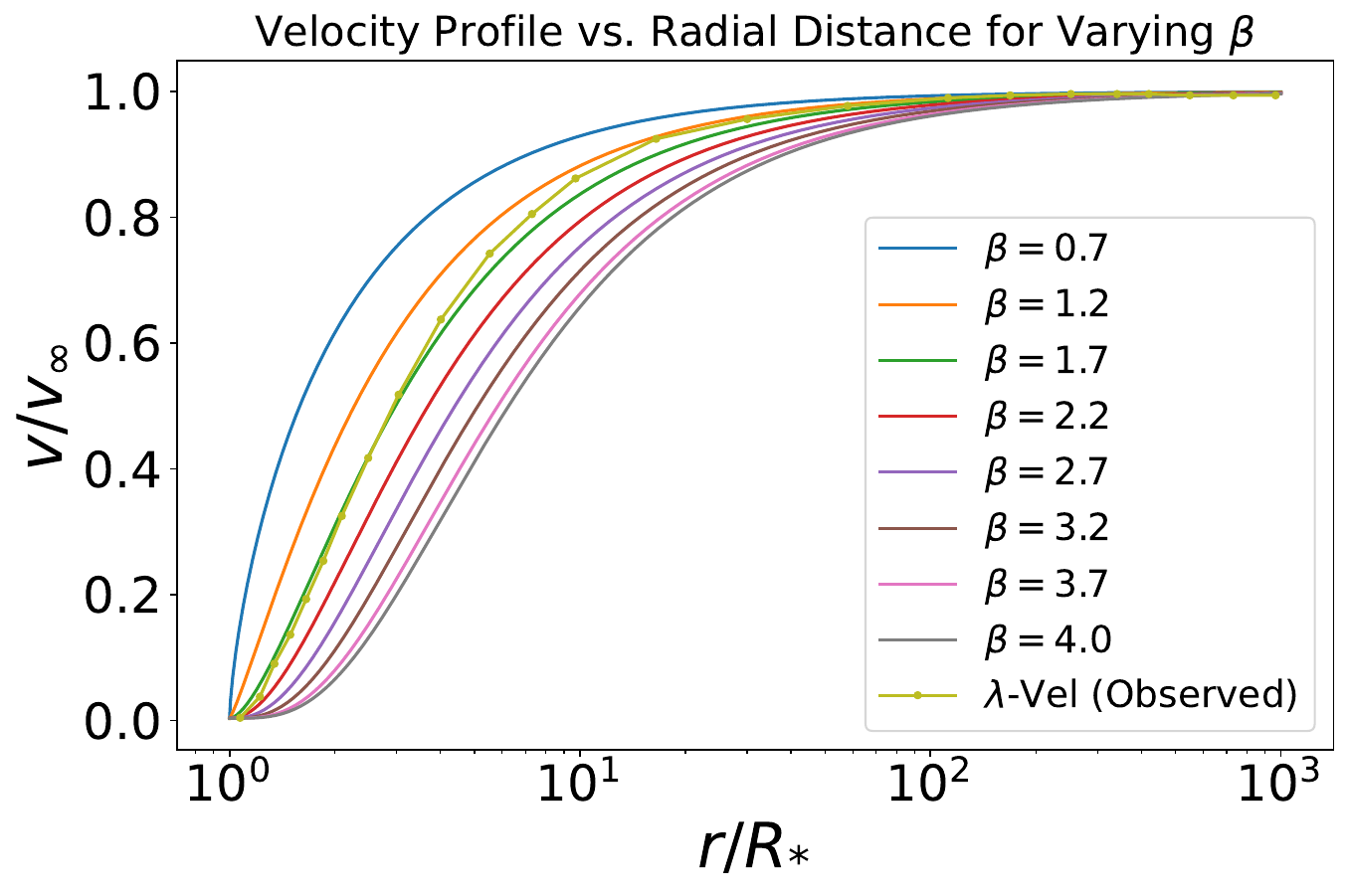}
    \caption{The radial variation of the wind velocity, $v(r)$, normalized to the terminal wind speed, $v_\infty=30~\mathrm{km~s}^{-1}$ of the RSG $\lambda$-Veloris, as per the $\beta$-law given by Equation~\eqref{equation : 4} for $\beta=0.7-4$, with $v_o=0.1~\mathrm{km~s}^{-1}~\mathrm{and~using}~R_*=1.46\times10^{13}~\mathrm{cm}$ for the radius of $\lambda$-Veloris \citep{Rau_2018}, with its observed velocity profile given by the dotted gray line \citep{2010ASPC..425..181B}.}
    \label{fig:figure11a}
\end{figure}

As described in Section~\ref{subsubsection : wind propagation}, due to lack of a better understanding of the driving mechanism for the RSG wind out to larger distances $\sim$~few stellar radii where dust can form to drag the gas along, we choose to parameterize this kinematic process with the $\beta$-velocity law as given by Equation~\eqref{equation : 4}. The normalized velocity profiles using this velocity law for various choices of $\beta$ commonly used in the literature is shown in Figure~\ref{fig:figure11a} which illustrates the change in the shape of the velocity profile, from convex to concave, very close to the stellar surface i.e. at $r\simeq R_*$, with increasing values of $\beta=0.7-4$. Comparing the shape of the velocity curves with that of observations of $\lambda$-Veloris \citep{2010ASPC..425..181B}, we select $\beta=1.2$ for the purpose of this study, as lower values of  $\beta$ result in steeper velocity profiles inconsistent with the observed values, while higher values of $\beta$ do not match the shape of the observed velocity profiles close to the stellar surface.

\section{Uncertainty estimates for calculating CSM densities}
The uncertainties in the values of CSM densities inferred from X-ray mass loss rates and terminal wind velocities of SNe 2023ixf \citep{2025arXiv250811747J} and 2020ywx \citep{Baer-Way_2025}, according to Equation~\eqref{equation : 5}, have been calculated by propagating the individual uncertainties in the estimation of these two properties from measurements of X-ray fluxes and fitting of spectral features, respectively. There are also additional uncertainties that involve assumptions regarding degree of ionization in the shock propagating through the CSM post explosion as well as the composition of the CSM but a quantitative estimate of these uncertainties are beyond the scope of this work.


\bibliographystyle{aasjournalv7}
\bibliography{Bibliography_sarangi,bibliograhy}

\end{document}